\documentclass[fleqn,usenatbib]{mnras}

\usepackage[T1]{fontenc}

\DeclareRobustCommand{\VAN}[3]{#2}
\let\VANthebibliography\thebibliography
\def\thebibliography{\DeclareRobustCommand{\VAN}[3]{##3}\VANthebibliography}

\usepackage{graphicx}   
\usepackage{tabularx}
\usepackage{amsmath}    
\usepackage{amssymb}    
\usepackage{orcidlink}
\usepackage{newtxtext,newtxmath}

\usepackage{xcolor}
\definecolor{pink}{rgb}{0.96, 0.76, 0.76}
\definecolor{aqua}{rgb}{0.22, 0.96, 0.93}
\definecolor{darkred}{rgb}{0.76, 0.23, 0.13}

\usepackage{color,soul}
\definecolor{lightblue}{rgb}{.70,.95,1}
\sethlcolor{lightblue}

\title[VMP Milky Way]{On the existence of a very metal-poor disc in the Milky Way}

\author[Zhang et al.]{
Hanyuan Zhang$^{1}$\thanks{E-mail: hz420@cam.ac.uk}\orcidlink{0009-0005-6898-0927},
Anke Ardern-Arentsen$^{1}$\orcidlink{0000-0002-0544-2217} and Vasily Belokurov$^{1}$\orcidlink{0000-0002-0038-9584}  \\
\\
$^{1}$ Institute of Astronomy, University of Cambridge, Madingley Road, Cambridge CB3 0HA, UK
}
\date{Accepted 2024 August 01. Received 2024 July 26; in original form 2023 November 15}

\pubyear{2023}

\begin{document}
\label{firstpage}
\pagerange{\pageref{firstpage}--\pageref{lastpage}}
\maketitle

\begin{abstract}
The question of whether the Milky Way's disc extends to low metallicity has been the subject of debate for many years. We aim to address the question by employing a large sample of giant stars with radial velocities and homogeneous metallicities based on the Gaia DR3 XP spectra. We study the 3D velocity distribution of stars in various metallicity ranges, including the very-metal poor regime (VMP, [M/H] $<-2.0$). We find that a clear, standalone disc population,  i.e. that with a ratio of rotational velocity to velocity dispersion $v/\sigma>1$, starts to emerge only around [M/H] $\sim -1.3$, and is not visible for [M/H] $<-1.6$. Using Gaussian Mixture Modeling (GMM), we show that there are two halo populations in the VMP regime: one stationary and one with a net prograde rotation of $\sim80\,\mathrm{km/s}$. In this low-metallicity range, we are able to place constraints on the contribution of a rotation-supported thick disc sub-population to a maximum of $\sim 3$\% in our sample. We compare our results to previous claims of discy VMP stars in both observations and simulations and find that having a prograde halo component could explain most of these.
\end{abstract}







\section{Introduction}

How and when do stable and dominant galactic stellar discs form? This question is currently being answered with high-redshift studies of galaxy morphology  \citep[see e.g.][]{Zhang_2019,Nelson_2023,Kartaltepe_2023,Robertson_2023,Ferreira_2023}, probes of gas kinematics \citep[see e.g.][]{Wisnioski_2019,Ubler_2019, Rizzo_2021, Fraternali_2021, deGraaff_2023,Pope_2023}, and detailed numerical simulations of galaxy evolution \citep[see e.g.][]{Stern_2021,Hafen_2022, Gurvich_2023, Hopkins_2023,Semenov_2023}. In the Milky Way (MW), chemo-kinematic analysis of ancient low-mass stars can provide strong, independent constraints on the emergence of the stellar disc \citep[see e.g.][]{Belokurov_2022,Conroy_2022,Xiang_2022}. As precise stellar ages are not yet readily available for old stars, stellar metallicity is routinely used instead as a proxy for a Galactic clock. 

To turn stellar abundance ratios into a clock, an age-metallicity relation needs to be established. Multiple attempts to do so observationally \citep[see e.g.][]{Nordstrom_2004,Haywood_2013,Sanders_2018, Queiroz_2018,Leung_2019,Feuillet_2019,Miglio_2021,Sahlhodt_2022,Anders_2023,Queiroz_2023, Wu_2023, Kordopatis_2023} agree that the MW's high-$\alpha$ disc stars with [Fe/H]\footnote{[X/Y] $ = \log(N_\mathrm{X}/N_\mathrm{Y})_* - \log(N_\mathrm{X}/N_\mathrm{Y})_{\odot}$, where the asterisk subscript refers to the considered star, and N is the number density for element X or Y. The denotations [Fe/H], ``metallicity'' and [M/H] are often used interchangeably, although technically they are not the same -- [Fe/H] is the iron abundance and [M/H] the overall metal abundance. However, in practice they are quite similar.}~ $\lessapprox-1$ have ages of $\gtrapprox10$ Gyr. To probe the earliest phases of the disc formation therefore requires the identification and characterisation of stars with metallicities below [Fe/H]~$=-1$. In this low-metallicity regime.

The bulk of the stars at low metallicity, at least those accessible to observations currently, likely formed elsewhere and were subsequently accreted onto the MW -- for example as part of the Gaia Sausage/Enceladus (GS/E) event \citep[see][]{Belokurov_2018,Haywood_2018,Helmi_2018,Deason_2018} and other, lower-mass mergers \citep[][]{Myeong_2018a,Myeong_2018b,Koppelman_2018}. The GS/E stars, characterised by a very high orbital anisotropy $\beta>0.8$, have been shown to contribute a large fraction of the inner accreted halo at $-2<$~[Fe/H]~$<-1$ \citep[see][]{Belokurov_2018,Deason_2018}. Above [Fe/H]~$=-1$, the stellar halo is dominated by the heated high-$\alpha$ disc, the population known as the Splash \citep[see][]{Bonaca_2017, Gallart_2019,Di_Matteo_2019,Belokurov_2020}. Below [Fe/H]~$\approx-2$, the contribution from smaller mass accretion events, distinct from the GS/E, grows substantially, bringing the overall halo anisotropy down \citep[see e.g.][]{Lancaster2019,Bird_2021}.

Despite the established general picture that most of the low-[Fe/H] stars must have been accreted, glimpses of the so-called metal-weak Galactic disc population, i.e. stars with $-2<\mathrm{[Fe/H]}<-1$ (or even below $-2$), apparent non-zero rotation (high $V_{\phi}$) and orbits with intermediate eccentricity, have loomed periodically in the observational literature \citep[e.g.][]{Norris_1985, Morrison_1990, Beers_1995, Chiba_2000, Carollo_2010, Ruchti_2011, Kordopatis_2013, Carollo_2019, AnBeers_2020} as well as in simulations \citep[e.g.][]{Sotillo-Ramos_2023}. The astrometry from the \textit{Gaia} mission \citep{gaiamission} has also made it possible to derive detailed orbital properties for large samples of the most metal-poor stars, and there have been numerous identifications of discy stars with [Fe/H]~$<-2.0$ and even down to $<-4.0$ \citep[e.g.][]{Sestito_2019, Sestito_2020, Venn_2020, Di_Matteo_2020, FernandezAlvar_2021, Cordoni_2021, Mardini_2022, Feltzing2023}, with a mix of interpretations among authors. 

\cite{Sestito_2019, Sestito_2020} discussed three possible origins of the very metal-poor (VMP, [Fe/H]~$<-2.0$) disc-like/planar stars: (1) they formed \textit{in-situ} in an early galactic disc; (2) they were born in the gas-rich building blocks of the proto-Milky Way, which formed the backbone of the later disc; (3) they are accreted from prograde minor mergers (after the disc is already in place). In the second scenario, the population can be a combination of stars born in smaller galaxies/building blocks and stars born in the main MW progenitor. In the first and second scenarios, the stars may need to be brought out from the inner Galaxy to the Solar radius (where we observe them) through radial migration and/or interactions with the Galactic bar. The origin of metal-poor disc-like/planar stars was further explored via the analysis of high-resolution cosmological simulations in e.g. \citet{Sestito_2021} and \citet{Santistevan_2021}, finding that simulations of MW-like galaxies also show an overdensity of planar very metal-poor stars. They conclude that the main contributors to the metal-poor prograde planar population are the early Galactic building blocks (large and small, scenario 2) and the later minor accretions (scenarios 3), but that the formation of the disc tends to happen after these very metal-poor stars formed, which is evidence against the first scenario. 

The recent observational findings of \citet{Belokurov_2022} are also evidence against the very early disc scenario -- they show that already below [Fe/H]~$=-1$ ordered stellar rotation starts to disappear among MW stars, at least within the in-situ population. \citet{Belokurov_2022} use {\it Gaia} astrometry and APOGEE spectroscopy to demonstrate that the observed MW in-situ stellar population with [Fe/H]~$<-1.3$ is kinematically hot, namely has a 1D rotational velocity dispersion of order of $\approx100$ km/s. At the same time, the net rotational velocity is low for this population ($V_\phi \lessapprox 50$~km/s, \citealt{Belokurov_2022, BK_insitu_clusters2023}). This pre-disc MW in-situ population, dubbed {\it Aurora}, is isolated by \citet{Belokurov_2022} to have $-2.0\lesssim$~[Fe/H]~$\lesssim-1.3$ in the vicinity of the Sun, but the authors surmise that it should extend to lower metallicities and its density should increase towards the Galactic centre. Indeed, recently, \citet{Rix_2022} demonstrated the existence of a large concentration of metal-poor stars ([M/H]~$< -1.5$) towards the Galactic centre, in agreement with the earlier analysis of the very metal-poor component of the Galactic bulge region \citep[e.g.][]{Arentsen_PIGS_I_2020,Arentsen_PIGS_II_2020}. The exact connection between Aurora and the metal-poor stars inside the Solar radius has not been established yet (see however \citealt{Belokurov2023nitrogen} for evidence for the Aurora density following a steep power-law with Galactocentric radius). However, if there is a very metal-poor, centrally concentrated tail to Aurora, then the combination of the broad distribution with a non-zero net prograde spin can create an excess of stars with high rotational velocity $V_{\phi}$. This suggestion fits with the second scenario of \citet{Sestito_2019, Sestito_2020}, where the rotation comes from the early building blocks of the proto-MW, including the main MW progenitor (Aurora). 

Other numerical simulations of MW-like galaxies also do not support the scenario in which the galactic disc forms early at very high redshift. As discussed in \citet{Belokurov_2022}, stars formed inside the main progenitor at high redshift, e.g. metallicities below [Fe/H]~$\approx-1.5$ corresponding to $z >2$ \citep[but likely even $z>4$, as discussed in][]{BK_insitu_clusters2023}, are originally in a messy and disturbed state. Subsequently, over the lifetime of the galaxy, these ancient stars phase-mix to be found today in a fattened, spheroidal and centrally-concentrated distribution. The morphology and the kinematics of a galaxy change significantly when it becomes massive enough to finally be able to form a prominent and stable disc \citep[e.g.][]{Hafen_2022, Stern_2021, Hopkins_2023}. 
However, as pointed out by \citet{Belokurov_2022}, in most simulations, MW-like systems typically experience the spin-up transition from the pre-disc state to a disc at metallicities significantly higher compared to what is observed in our Galaxy, suggesting that the MW may have had an early disc formation. 
While some of this inconsistency may be due to the sub-grid recipes utilised, there are also genuine differences between the observed MW and the bulk of the MW-mass galaxies simulated. This has been recently explored by \citet{Semenov_2023} and \citet{Dillamore_2023spin} who show that in a small subset of models (of order of $\approx10\%$), a disc spin-up at metallicities consistent with the MW observations is possible if the host galaxy has a mass assembly which peaks early.

\begin{figure*}
    \centering
    \includegraphics[width = \textwidth]{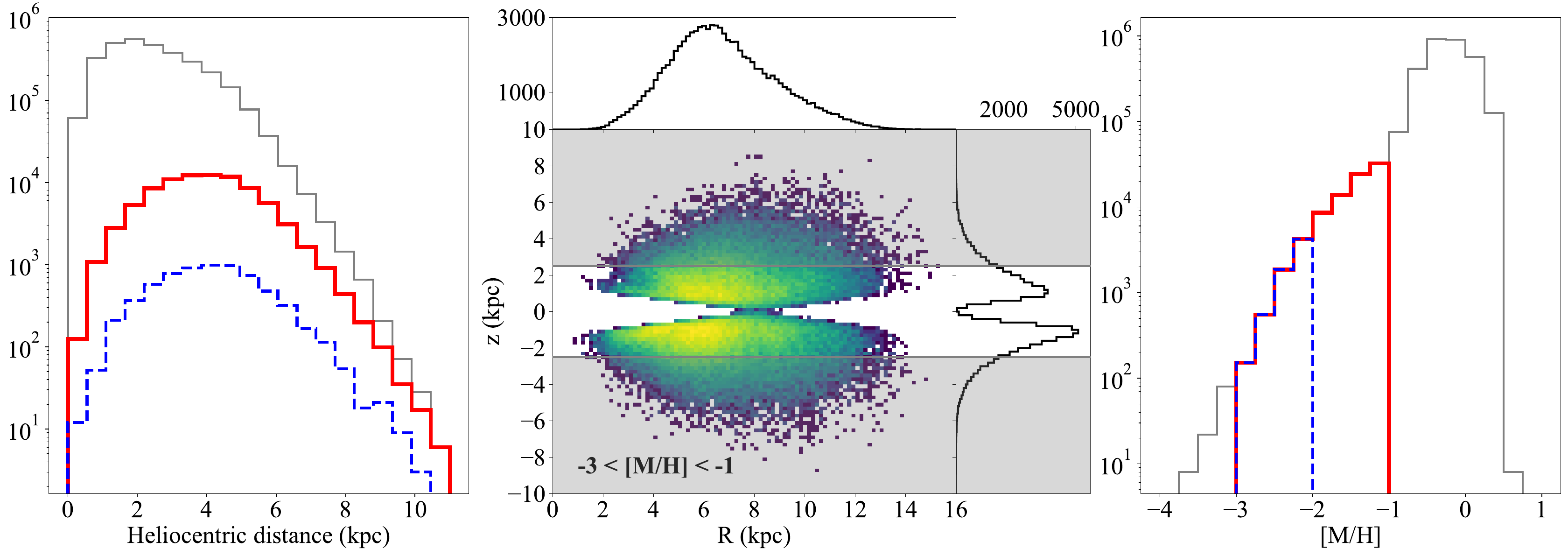}
    \caption{Properties of the final data sample after all selection cuts. Left: the heliocentric distance distribution. The solid red and dashed blue lines represent the distribution of stars with $-3<\mathrm{[M/H]}<-1$ and $-3<\mathrm{[M/H]}<-2$ respectively, while the grey line shows the distribution of the whole sample. Middle: the $R-z$ distribution for stars with $-3<\mathrm{[M/H]}<-1$, in which the area at $|z|>2.5$~kpc is greyed out as we mainly focus on the analysis of stars with $|z|<2.5$~kpc. Right: metallicity distribution of our data sample. Line designation is the same as in the left panel.}
    \label{fig:Sample_info}
\end{figure*}

Finally, even if no net spin is present to begin with in the population of stars on halo-like orbits, over time, a prograde-retrograde asymmetry can be created by interactions with a rotating Galactic bar \citep[e.g.][]{PerezVillegas_2017, Dillamore_2023bar}. \citet{Dillamore_2023bar} show that a noticeable number of halo stars can get trapped in resonances with the growing or slowing down bar. In their simulation, halo stars trapped in the bar co-rotation resonance create a pile-up at the total energy level matching the orbital frequency corresponding to the bar pattern speed. Even though the model distribution of stars in the $J_{\phi}-J_z$ action space is symmetric at the start of the simulation, an overdensity in $J_\phi-J_z$ space similar to that observed by \citet{Sestito_2019,Sestito_2020} develops over time. \citet{Dillamore_2023bar} also show that a very similar overdensity of stars is observed in the {\it Gaia} RVS data for a model with a realistic bar pattern speed, suggesting that bar-halo interactions have been affecting the kinematics of the MW low-metallicity stars. Note that according to \citet{Dillamore_2023bar}, the prograde bias the bar induces in the halo population is a strong decreasing function of Galactocentric distance (see their Fig.~10), and is therefore less strong locally compared to the inner Galaxy. 

In this work, we study the kinematic properties of a large homogeneous sample of stars with radial velocities from the RVS sample of the {\it Gaia} Data Release 3 \citep[DR3,][]{Gaia_DR3} and metallicities based on Gaia DR3 XP spectra from \citet{Andrae_2023}. Our sample covers a wide range in metallicity ($-3.0 <$ [M/H] $< +0.5$). 
We focus on the behaviour of the most metal-poor stars, and investigate whether there appears to be a significant fraction of very metal-poor stars that can be associated with the Galactic disc. In short, in our sample, we find that the contribution from the thick disc to the VMP population can be constrained to be less than 3\%
(even after limiting the sample $|z| < 2.5$~kpc, where $z$ is the distance away from the Galactic plane). We suggest that the overdensity of prograde VMP stars is due to a prograde halo component and not due to the presence of a significant disc population. With the present sample, we cannot exclude the presence of a minority population of VMP stars on disc orbits, especially in the thin disc regime.

We present the metallicity sample used for our investigation in Section~\ref{sec::data}, as well as the selection cuts we made to construct our final sample. In Section~\ref{sec::model}, we present the results of a Gaussian mixture model (GMM) analysis of the 3D velocity distributions in various metallicity ranges, including the VMP regime. We discuss the limitations of our approach in Section~\ref{sec::discussion}, and put our results into context by comparing them with previous claims for the existence of a very metal-poor disc. In Section~\ref{sec::simulation}, we put recent perspectives on disc formation in simulations of Milky Way-analogue galaxies together into a coherent scenario. We summarise our conclusions in Section~\ref{sec::conclusions}.

\section{Data}\label{sec::data}

\subsection{XGBoost metallicity sample}
\label{subsec::XGBoost_sample}

\cite{Andrae_2023} derived stellar parameters of 175 million stars with \textit{Gaia} DR3 XP spectra using the XGBoost algorithm. Their training sample consists of stars from APOGEE DR17 \citep{APOGEE_DR17} and an additional very/extremely metal-poor star sample from \cite{Li_2022_metalpoorstar}. Inputting various broadband and narrow-band photometric measurements (ranging from optical to infrared), the \textit{Gaia} XP spectrum coefficients and parallax, the trained XGBoost model predicts the metallicity [M/H], effective temperature, and surface gravity.
\cite{Andrae_2023} claim that the uncertainty is $0.1$~dex in metallicity, $50$~K in effective temperature, $0.08$ dex in surface gravity. As expected, the precision and accuracy drop for fainter objects (and it is also expected to drop for more metal-poor stars). Here, we only use their vetted bright ($G < 16$) red-giant branch (RGB) star sample \citep[Table~2 in ][]{Andrae_2023}, which contains 17,558,141 stars with high-confidence metallicity in the range from $\mathrm{[M/H]}\sim-3$ dex to beyond solar metallicity. This RGB star sample was selected according to $G$ band apparent magnitude, fractional parallax uncertainty (fpu), and effective temperature and surface gravity derived from the XGBoost algorithm (the details are listed in section 4.2 of \citealt{Andrae_2023}). The sample is cross-validated against various surveys (e.g. GALAH, GSP-Spec, and SkyMapper) and demonstrated good consistency and accuracy, even at very low metallicities \citep[see also][]{Martin_2023_Pristine}. Comparing the metallicity $\mathrm{[M/H]}$ in \citet{Andrae_2023} to the iron-abundance $\mathrm{[Fe/H]}$ in APOGEE and GALAH shows a small offset on the order of $0.1-0.15$ dex, with the differences being somewhat larger for alpha-poor stars. The small offset would not affect the discussion in this work as we make no strong claims with regard to exact metallicity boundaries. 

\subsection{Sample construction}\label{subsec::build_sample}

\textit{Gaia} XP spectra are affected by Galactic dust through reddening and extinction. Dust extinction makes objects dimmer and thus, at fixed exposure time, reduces the signal-to-noise achieved. As result, stars in the high extinction and low galactic latitude regions are relatively under-represented in the XGboost metallicity sample. To avoid a complicated selection functions and assure a higher quality sample, we manually introduce two further cuts by removing stars with $\texttt{E(B-V)}_\texttt{{SFD}}>0.5$ or $|b|<10^\circ$, where $\mathrm{E(B-V)}$ is the colour excess from the SFD dust map \cite{SFD} queried using the \textsc{dustmaps} package \citep{dustmaps}, and $b$ is the Galactic latitude. We also remove stars within $1\,\mathrm{deg^2}$ of known Galactic globular clusters and dwarf galaxies. 

\begin{figure*}
    \centering
    \includegraphics[width = \textwidth]{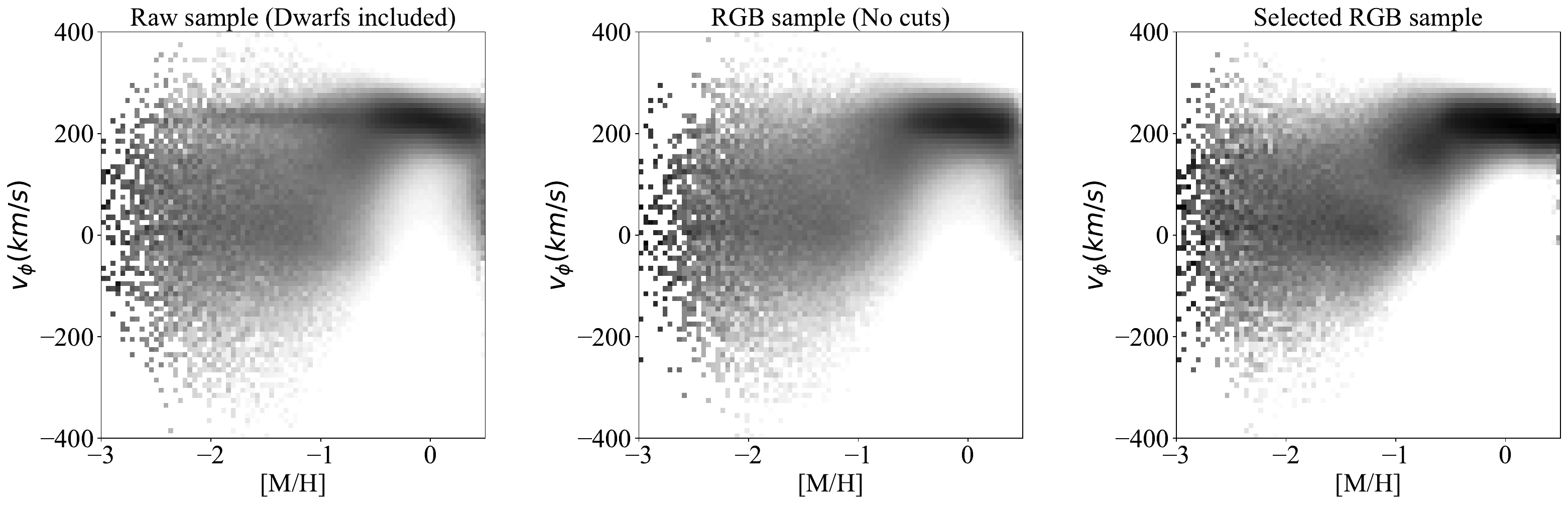}
    \caption{Column-normalised 2D histogram of stars in the [M/H]-$v_\phi$ plane.  Left panel shows the raw sample that includes all stars in \citet{Andrae_2023} without any cuts. A narrow band of fast-rotating stars can be seen extending into [M/H]$<-1$. This spurious signal is produced by hot stars with incorrectly assigned metallicities. Middle panel is for stars from the RGB sample with vetted metallicity in \citet{Andrae_2023} before we further apply our selection cuts. To plot the left and middle panels, we randomly select stars from the sample so that the total number of stars is roughly the same as in our final sample.  Right panel displays our final selected sample and reveals the sharp transition at [M/H]~$\sim -1.3$ that marks the spin-up phase of the Milky Way, intuitively corresponding to the epoch of the stellar disc formation. }
    \label{fig:feh_vphi}
\end{figure*}

The astrometric measurements and the radial velocity measurements are from \textit{Gaia} DR3 \citep{gaiamission,Gaia_DR3}, and we use the photo-geometric distance provided by \cite{BJ_2021} instead of 1/parallax. The distance estimate is crucial for obtaining accurate kinematics. The photo-geometric distance estimation is more accurate when the fpu is small; hence, we select stars with $\mathrm{fpu}<0.1$. We apply a strict fpu cut because i) we require high-quality kinematic measurements for this investigation, and ii) the sample size is large enough to make a statistical conclusion even after this cut. 

The orbital parameters are calculated using the Milky Way potential from \cite{McMillan_2017}. We performed numerical orbital integration for each star with a step-size of $1\,\mathrm{Myr}$ for $3\,\mathrm{Gyr}$ using \textsc{galpy} \citep{Bovy_2015}. The maximum height above or below the disc plane, $z_{max}$, and the orbital eccentricity, $e$, are subsequently obtained. The energy, $E$, of each orbit is also obtained using \textsc{galpy}. The actions ($J_r$ for radial action, $J_z$ for vertical action, and $J_\phi$ (or $L_z$) for azimuthal action) are calculated using \textsc{Agama} \citep{Vasiliev_2019} with the St\"ackel fudge method \citep{Binney_2012, Sanders_2016}.

We present the distribution of relevant parameters for our final sample in Fig.~\ref{fig:Sample_info}. 
The heliocentric distance and metallicity distributions are shown in the left and right panels, respectively. The solid red lines indicate the distributions of our metal-poor sample ($-3.0 <$~[M/H]~$<-1.0$) and the dashed blue lines are for very metal-poor stars only ($-3.0 <$~[M/H]~$<-2.0$). The heliocentric distances are skewed to larger values for metal-poor stars compared to the full sample. The distribution of Galactocentric cylindrical $R$ versus $z$ is shown in the middle panel, for the metal-poor sample only. The distribution of R reaches as close as 2~kpc and as far as 14~kpc from the Galactic centre, while there is an overdensity towards the inner Galaxy (low R). 

There are many stars in our sample within the expected spatial extent of the MW thick disc component (scale height $\sim 1.0$~kpc, \citealt{Bland-Hawthorn_2016}). However, outside of the immediate Solar neighbourhood, the thin disc region  (scale height $\sim 0.3$~kpc, \citealt{Bland-Hawthorn_2016}) is mostly excluded from our footprint. As we demonstrate in following sections, while thin (kinematically-cold and metal-rich) disc stars are present in our sample, by construction our sample is most appropriate for investigating the presence of a very metal-poor \textit{thick} disc component. Throughout this work, we will use the short-hand ``disc'' for any disc population, keeping in mind this is mostly probing the thick disc.

It is important to select a high quality sample for our analysis. For comparison, in Fig.~\ref{fig:feh_vphi} we present the column-normalised 2D histogram in the $\mathrm{[M/H]}$-$v_\phi$ plane for stars with \textit{Gaia} radial velocities that are in the raw \citet{Andrae_2023} sample without any further selection (left), for their sample of vetted RGB stars before we apply additional cuts (middle), and our carefully selected sub-sample of RGB stars after cuts on E(B$-$V), $|b|$, fpu and having removed substructures (right). The raw sample also contains turn-off and dwarf stars, for which the metallicities are expected to be less reliable \citep{Andrae_2023}, especially for metal-poor stars because there are not many metal-poor dwarfs in the training sample and/or the spectral features become weaker due to the higher stellar temperatures for turn-off stars. A conspicuously sharp overdensity of stars with high rotation at low metallicity can be seen in the raw sample (left panel). A cross-match between the \citet{Andrae_2023} sample and LAMOST DR8 \citep{Deng_2012_LAMOST, LAMOST_Xiang_2015, LAMOST_Xiang_2017} reveals that a significant number of hot metal-rich dwarfs ([Fe/H]~$\gtrsim 0$, mostly $T_{\mathrm{eff}} > 9000$~K, but also some turn-off stars with $T_{\mathrm{eff}} \sim 6000-7000$~K) are assigned low XGBoost metallicity. The ``thin disc sequence'' in the left-hand panel is therefore likely due to metal-rich contamination (see more details in Appendix~\ref{Appendix::0}). The contamination can be easily removed when using only the vetted RGB stars, as shown in the middle panel of Fig.~\ref{fig:feh_vphi}. 


\section{Chemo-kinematic decomposition of the Milky Way}\label{sec::model}

\begin{figure*}
    \centering
    \includegraphics[width = \textwidth]{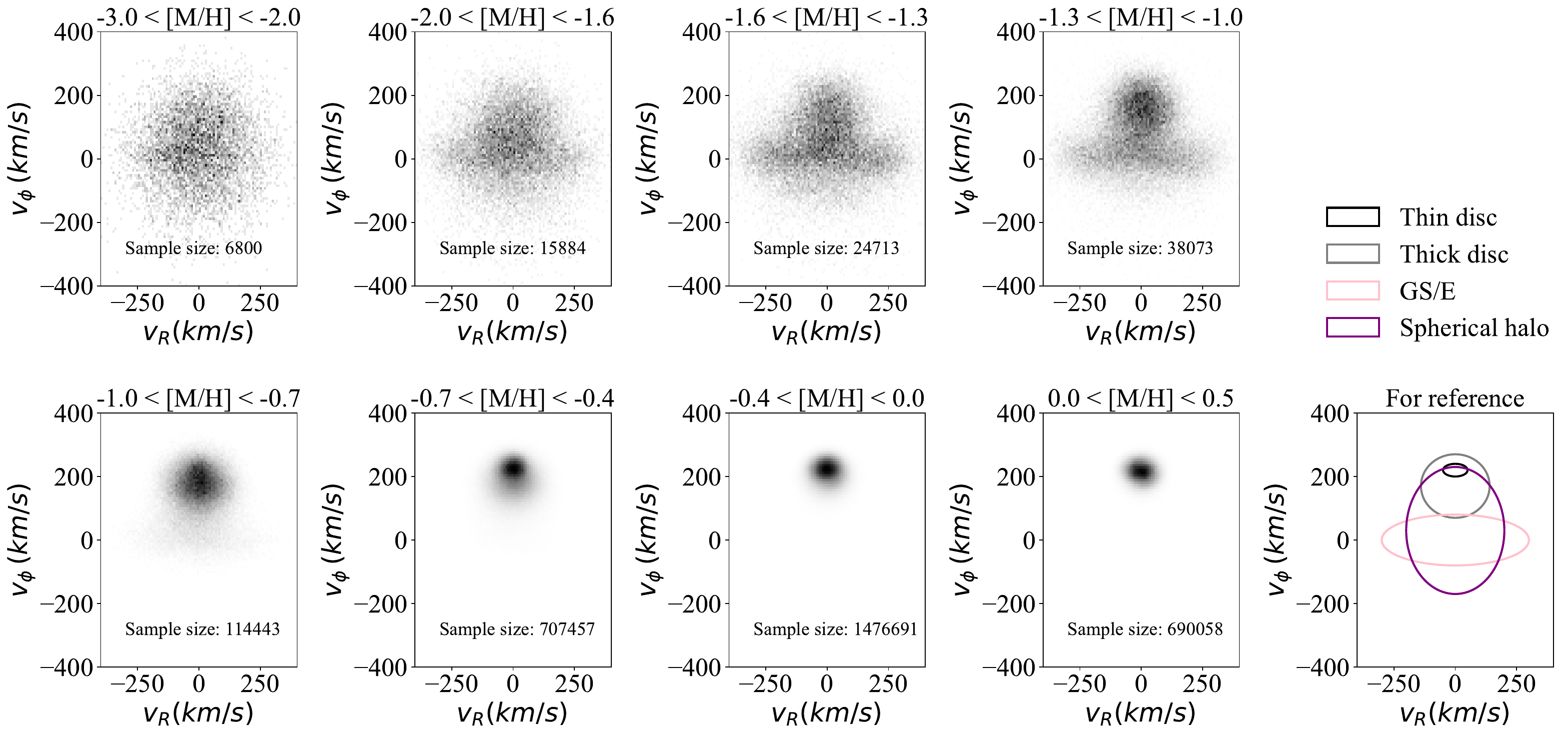}
    \caption{This shows the $v_R$-$v_{\phi}$ distribution for the sample in various metallicity bins, in which the sample size in each bin is noted in each panel. The lower right panel is drawn for reference, showing the expected location for several established Milky Way components, including two rotation-supported discs, the pressure-supported halo as well as the GS/E. The GS/E component is visible over a wide span of metallicity $-2<$[M/H]$<-0.7$ but its contribution at either end of this range is noticeably reduced. The velocity distribution is the broadest at [M/H]$<-2$ and the narrowest at [M/H]$>0$.}
    \label{fig:vr_vphi_space}
\end{figure*}

Combining the metallicity predicted in \cite{Andrae_2023} and the kinematics measurements from \textit{Gaia} DR3, we track the chemo-kinematic evolution of the Milky Way by binning stars according to their metallicity in the range of $-3<\mathrm{[M/H]}<+0.5$ -- assuming the metallicity is correlated with Galactic time. Note that this assumption is only approximately correct for stars born in a single galaxy and does not strictly hold for the MW in-situ stars with [Fe/H]~$>-1$ where two distinct age distributions in high-$\alpha$ and low-$\alpha$ discs overlap in metallicity space. In the low-metallicity regime, we expect a contribution from the accreted stars formed outside of the MW. Below, we mainly focus the discussion on whether there is any strong evidence for a disc component among the very metal-poor stars.

\subsection{Distributions of radial and azimuthal velocities as function of [M/H]}\label{subsec::qualitative}

We first investigate the distribution of azimuthal velocities as function of metallicity in the right panel of Fig.~\ref{fig:feh_vphi}. The figure shows that there is a transition from rotation-dominated orbits (characterised by high azimuthal velocity $V_{\phi}$ and low velocity dispersion) above $\mathrm{[M/H]}\sim-1.0$ to pressure-supported orbits (slow or zero rotation, $v_\phi \approx 0$, and high velocity dispersion) at lower metallicities ([M/H]~$<-1.0$). Overall, at low metallicity, i.e. [M/H]$~<-1.5$, there appears to be a systematic prevalence of positive azimuthal velocities, but the clear disc sequence ($v_\phi \gtrsim 150$~km/s) does not extend below [M/H]~$\lesssim -1.3$. 

We further investigate the distributions of stars in the $v_R$--$v_{\phi}$ space as a function of metallicity bins in Fig.~\ref{fig:vr_vphi_space}, where $v_R$ is the galactocentric cylindrical radial velocity. In the bottom right corner of the Figure the expected locations of different Milky Way components are marked for illustration. Here the low-$\alpha$ (a.k.a. thin, black line) and the high-$\alpha$ (a.k.a. thick, grey line) discs have high rotation, small velocity dispersion and low $v_R$, while the halo components (pink and purple lines) have small rotation and large velocity dispersions in both radial and azimuthal directions. \textit{Gaia}-Sausage/Enceladus (GS/E) is dominated by radial orbits, with low net rotation and a large range of $v_R$. 

Fig.~\ref{fig:vr_vphi_space} demonstrates that in the VMP regime ($-3<\mathrm{[M/H]}<-2$), the velocity distribution is halo-like, i.e. approximately isotropic with little net rotation and without obvious disc-like features. We will place constraints on the disc fraction in Section~\ref{subsec::disc_fraction} using a more quantitative analysis. In the metal-rich bins ($\mathrm{[M/H]}>-0.7$), the thin disc population dominates the sample as expected. The sharpest transition from the halo-dominated era to the disc-dominated era, hence, happens around $-2<\mathrm{[M/H]}\leq-0.7$ when the behaviour in the $v_R$-$v_{\phi}$ space changes rapidly among the four metallicity bins. Visually, the disc signature disappears in the metallicity bin of $-2<\mathrm{[M/H]}\leq-1.6$; the thick disc quickly forms during the time corresponding to metallicities of $-1.3<\mathrm{[M/H]}\leq-1$ and subsequently starts to dominate the sample in the next metallicity bin. At higher metallicities, changes in the azimuthal velocity with [M/H] are much less dramatic. \cite{Belokurov_2022} use stars from \textit{Gaia}-APOGEE and a number of numerical simulations to illustrate that the Milky Way spun up rapidly between metallicity of $-1.3<\mathrm{[M/H]}<-0.9$, which is in agreement with our observations here. Note however that \cite{Belokurov_2022} analysed in-situ stars only, while in our analysis we do not have chemical tags to make a distinction between the in-situ and accreted stellar populations. Instead, in what follows we decompose the velocity distribution into individual components as a function of metallicity.

\subsection{Gaussian mixture models of the velocity distribution}\label{subsec::BIC_analysis}

\begin{figure*}
    \centering
    \includegraphics[width = \textwidth]{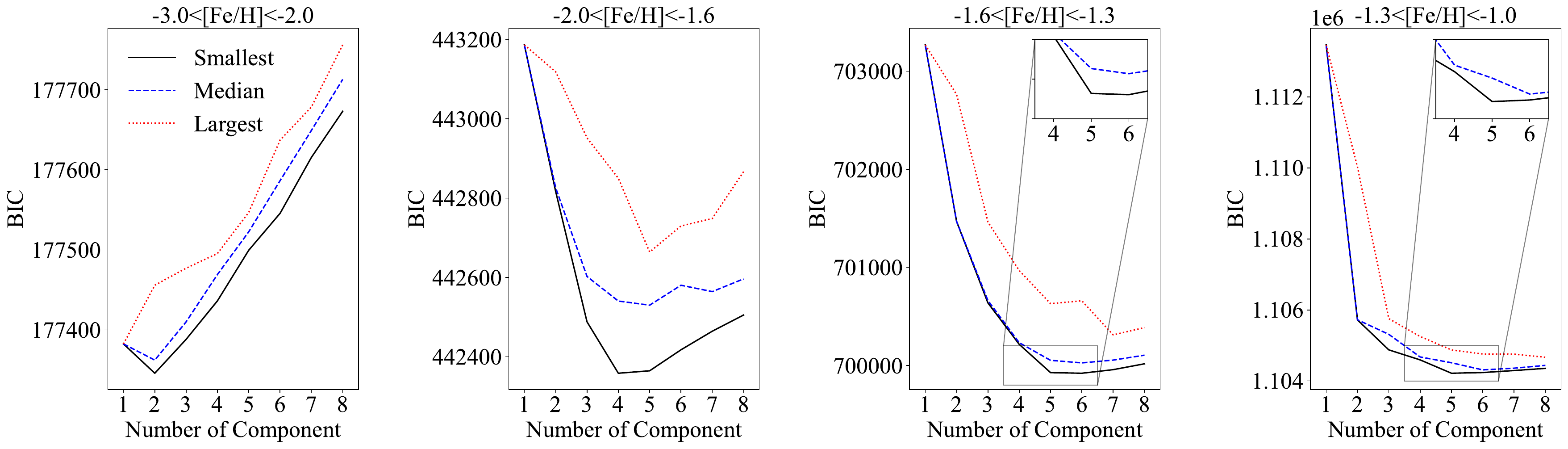}
    \caption{GMM's Bayesian information criteria (BIC) as a function of the number of model components in four metallicity bins between $-3<\mathrm{[M/H]}<-1$. In each [M/H] bin, we perform GMM fitting with different, random initial conditions 50 times, and the smallest, median, and largest BIC values are plotted in black-solid, blue-dashed, and red-dotted lines, respectively. The smallest BIC value indicates the optimum GMM fitting in that metallicity bin.}
    \label{fig:BIC}
\end{figure*}

\begin{figure*}
    \centering
    \includegraphics[width = \textwidth]{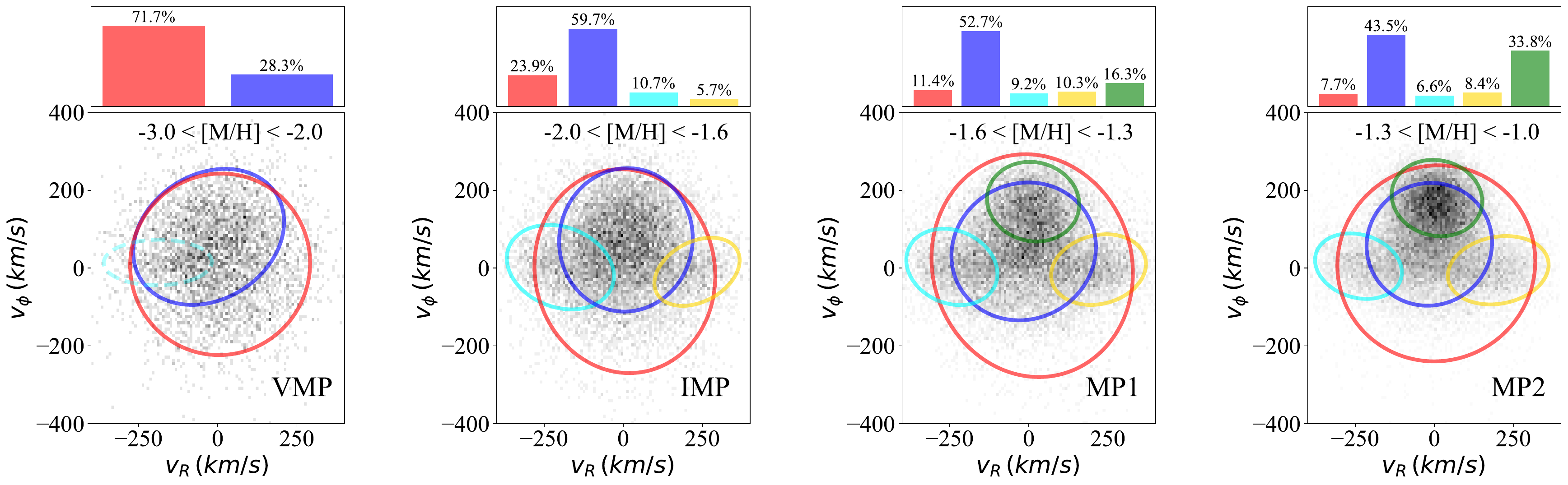}
    \caption{Optimum (as indicated by BIC) GMM for each metallicity bin, where each ellipse is the $2\sigma$ contour of the corresponing model Gaussian for each sub-population. On the top of each panel, the histogram gives the fractional contribution for each identified substructure. Across [M/H] bins colour remains the same for  Galactic components that are kinematically similar: red for the stationary halo, blue for the prograde halo, aqua and gold for GS/E components, and green for the rotation-supported disc. In the leftmost panel, the VMP bin, the aqua-dashed ellipse is not shown as the corresponding component is not identified in the optimum two-component GMM model, but it is recognised in the three-component GMM fit. 
    The possible existence of this minor substructure can explain the tilting of the blue ellipse in the VMP regime and the asymmetry of the GS/E components in the IMP bin. }
    \label{fig:Best_fit_GMM}
\end{figure*}

To decipher the early structures of the Milky Way, we employ the Gaussian mixture models (GMM) to help with the quantitative analysis. The Gaussian mixture model is an unsupervised learning algorithm that treats the distribution of each sub-population in a sample as an N-dimensional Gaussian distribution. The model is described by i) the weighting factor (the fractional contribution), ii) the mean, and iii) the covariance of of each Gaussian sub-population. We use \textsc{pyGMMis} \citep{pyGMMis}, which fits GMMs to data using the Expectation-Maximisation algorithm with the ``Extreme Deconvolution'' techniques developed by \cite{XDbovy}. The advantage of the Extreme Deconvolution approach is that it accounts for the measurement uncertainty when performing GMM fitting.

We concentrate our analysis on metal-poor stars ($-3<\mathrm{[M/H]}<-1$). We do not attempt to fit GMMs to more metal-rich bins, partly because the distribution of the thin disc stars deviates significantly from a Gaussian-like distribution in $v_R-v_{\phi}$ space, so the accuracy and validity of fitting are not guaranteed. To help reveal the disc population possibly contained within our sample, we further remove stars with $|z|>2.5$ kpc. In each metallicity bin, the GMM is produced in the space spanned by the three velocity components in the galactocentric cylindrical coordinate system, $v_R$, $v_\phi$, and $v_z$. The input error ellipse for each star is a diagonal covariance matrix with the square of uncertainties in $v_R$, $v_\phi$, and $v_z$ at the respective locations.

A common problem for the GMM analysis is finding the balance between overfitting and underfitting. Following the conventional approach \citep[see e.g.][]{Myeong_2022}, we choose the appropriate number of Gaussian components according to the Bayesian information criteria \citep[BIC,][]{BIC}. We calculate BIC using
\begin{equation}
    \mathrm{BIC} = k\ln(n)-2\ln L,
    \label{BIC_eq}
\end{equation}
where $k=(1+3+6)\times N-1$, is the total number of model parameters for $N$ Gaussian components, $n$ is the size of the sample, $\ln L$ is the log-likelihood of the data. Because a local minimum can easily trap the GMM, we run the GMM fit with different random initialisation 50 times and record the BIC value for each trial. The fit with the lowest BIC value indicates the optimal case, indicating that the global minimum was likely reached. In Fig.~\ref{fig:BIC}, we plot the BIC values in four metallicity bins as a function of $N$, and colour the $0$th (smallest), $50$th (median), and $100$th (largest) percentile of the BIC value for each $N$ by black, blue, and red lines, respectively. Hence, inspecting Fig.~\ref{fig:BIC}, we find the preferred component numbers are $N = 2$, $4$, $5$, and $5$ for $-3.0<\mathrm{[M/H]}<-2.0$ (very metal-poor, VMP), $-2.0<\mathrm{[M/H]}<-1.6$ (intermediate metal-poor, IMP), $-1.6<\mathrm{[M/H]}<-1.3$ (metal poor 1, MP1), and $-1.3<\mathrm{[M/H]}<-1.0$ (metal-poor 2, MP2) bins, respectively, based on the black lines. 

Increasing the complexity of the model with an additional component while keeping the log-likelihood invariant increases the BIC value by order of $10\times\ln(n)\sim100$. Comparing to this order of magnitude, some $N$-component GMMs share very similar BIC values (e.g. in the IMP bin, the BIC values of the 4-component model and the 5-component model only differ by $6.2$; in the MP1 bin, the BIC values of the 5-component model and the 6-component model only differ by $6.1$). We prefer the models with fewer components as the optimised fitting when BIC values are similar because they have a more straightforward physical interpretation. We discuss this further in Appendix~\ref{Appendix::A}, where we show the full GMM fitting result in Fig.~\ref{fig:appendix}. The additional components are generally added only to complicate the halo structure, and are not not in the disc area of the $v_R$--$v_\phi$ space. We therefore conclude that our results regarding the presence of disc components are robust.  

Fig.~\ref{fig:Best_fit_GMM} presents the best-fit GMMs in $v_R$--$v_\phi$ space for the preferred number of components in each metallicity bin (2 for VMP, 4 for IMP, 5 for MP1 and MP2). Here, each coloured ellipse represents the model Gaussian component of each sub-population. We interpret these components as a stationary halo, a prograde halo, GS/E (in two parts) and the thick disc. The parameters of the best-fit GMMs are given in Table~\ref{table:GMM_params}. We compute the uncertainty of the GMM parameters by re-generating the $v_R$, $v_{\phi}$, and $v_z$ according to the measurement error for each individual star and repeat the GMM fitting using the previous procedure 100 times. The uncertainty of the GMM parameters is on the order of $0.1$~km/s in general.

\begin{table*}
\caption{Parameters of the Gaussian mixture model fittings in different metallicity bins. The unit for columns of velocity is km/s. }
\begin{center}
\begin{tabular}{lccccccc}
\hline
\hline
 Components  & Weights ($\%$) & $\widebar{v_R}$ & $\sigma_R$   & $\widebar{v_\phi}$ & $\sigma_{\phi}$  & $\widebar{v_z}$ & $\sigma_{z}$\\
\hline
\multicolumn{8}{l}{\textbf{VMP}: $-3.0<\mathrm{[M/H]}<-2.0$ (4772 stars)}\\
\hline
Stationary halo & 71.7 & 7.28 & 141.8 & 9.6 & 116.6 & -1.2 & 116.2\\
Prograde halo & 28.3 & -28.7 & 118.9 & 80.0 & 87.4 & 3.15 & 69.6\\
\hline
\multicolumn{8}{l}{\textbf{IMP}: $-2.0<\mathrm{[M/H]}<-1.6$ (12062 stars)}\\
\hline
Stationary halo & 23.9 & 4.9 & 142.6 & -8.0 & 131.0 & -1.2 & 123.0 \\
Prograde halo & 59.7 & 8.1 & 105.4 & 72.4 & 92.2 & -0.7 & 72.0 \\
GS/E(1) & 5.7 & 232.5 & 67.3 & -9.6 & 43.8 & -7.0 & 89.9\\
GS/E(2) & 10.7 & -198.8 & 84.5 & 2.3 & 54.3 & 8.4 & 85.6\\
\hline
\multicolumn{8}{l}{\textbf{MP1}: $-1.6<\mathrm{[M/H]}<-1.3$ (19176 stars)}\\
\hline
Stationary halo & 11.4 & 10.5 & 158.9 & 6.3 & 143.1 & -4.1 & 131.4\\
Prograde halo & 52.7 & -15.9 & 113.9 & 43.0 & 88.6 & 3.0 & 71.2\\
GS/E(1) & 10.3 & 220.2 & 74.8 & -3.9 & 45.3 & -2.2 & 91.2\\
GS/E(2) & 9.2 & -243.0 & 71.9 & 2.9 & 49.1 & 4.9 & 94.0\\
Thick disc  & 16.3 & 13.5 & 72.9 & 170.8 & 51.2 & -10.2 & 67.0\\
\hline
\multicolumn{8}{l}{\textbf{MP2}: $-1.3<\mathrm{[M/H]}<-1.0$ (30884 stars)}\\
\hline
Stationary halo & 7.7 & 6.1 & 156.2 & 12.0 & 126.0 & -4.2 & 115.2 \\
Prograde halo & 43.5 & -16.0 & 99.0 & 61.1 & 79.0 & -1.3 & 70.7\\
GS/E(1) & 8.4 & 201.5 & 79.5 & -5.7 & 44.0 & -3.7 & 88.4\\
GS/E(2) & 6.6 & -239.4 & 68.5 & 4.8 & 42.1 & 3.7 & 89.2\\
Thick disc & 33.8 & 8.4 & 71.4 & 180.0 & 49.1 & 1.0 & 61.1\\

\hline
\hline
\end{tabular}
\end{center}
\label{table:GMM_params}
\end{table*}

\subsection{Model GMM components as a function of [M/H]}\label{subsec::GMM_results}
\begin{figure}
    \centering
    \includegraphics[width = \columnwidth]{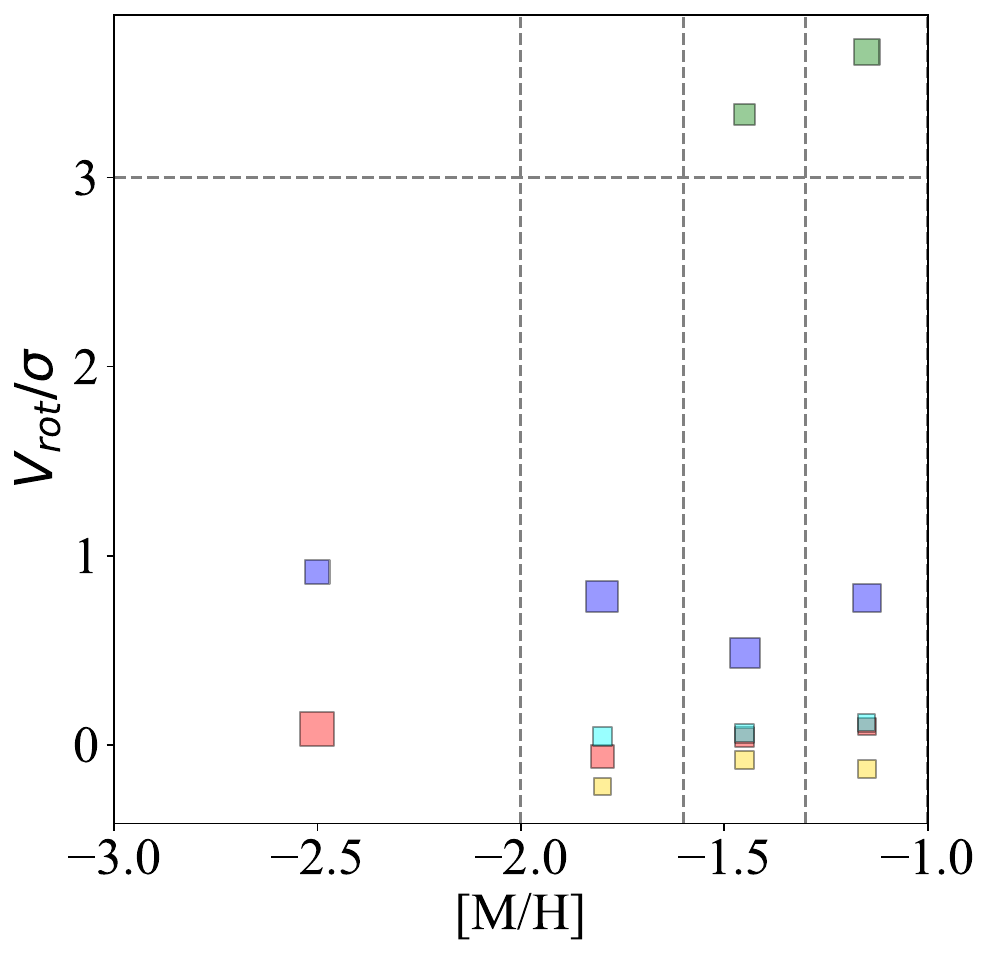}
    \caption{The ratio of the mean rotation velocity over the velocity dispersion in the azimuthal direction, $V/\sigma$ for individual Gaussian components identified in each metallicity bin. The colour-coding is the same as in Fig.~\ref{fig:Best_fit_GMM}. The size of the square represents the fractional contribution of each component. Horizontal dashed line at $V/\sigma=3$ shows the conventional boundary for the rotation-supported disc. The uncertainties are negligible, so they are not plotted in this diagram.}
    \label{fig:V_sigma}
\end{figure}

We continue our discussion from Sec.~\ref{subsec::qualitative} regarding the presence or absence of a disc sub-population with decreasing metallicity. Fig.~\ref{fig:V_sigma} shows the ratio of the rotational velocity to the azimuthal velocity dispersion, $V_{\mathrm{rot}}/\sigma$, for different GMM components in each metallicity bin. All identified structures consistently show little evolution of  $V_{\mathrm{rot}}/\sigma$ in the range of $-3<\mathrm{[M/H]}<-1$. No rotation-supported structure is recognised by the GMM in the VMP and IMP bins, but a rotation-supported, disc-like population with $V_{\mathrm{rot}}/\sigma > 3$, and $\widebar{v_{\phi}}\sim 170$~km/s is found in the two MP bins ([M/H]~$> -1.6$). The weight factor for the disc population increases from $16.3\%$ in the MP1 bin to $33.8\%$ in the MP2 bin, consistent with rapid disc growth in this metallicity range. Next, we discuss the other identified GMM components. 

The two components identified by the GMM for VMP stars (left panel in Fig.~\ref{fig:Best_fit_GMM}) both have velocity dispersions larger than their mean velocities, hinting at their halo-like nature. One of the components (red ellipse) does not have a significant net rotation ($\widebar{v}_{\phi}\approx0$~km/s) and a high velocity dispersion, hereafter we will refer to this as the stationary halo. In the VMP range, it corresponds to $\sim 70\%$ of the stars. The other (blue ellipse) has a net positive rotational velocity $\widebar{v}_{\phi} \approx 80$~km/s, and lower dispersion in $v_\phi$ -- hereafter we will refer to this as the prograde halo. These two components are also found in all other metallicity bins. The velocity dispersion ellipsoid of the prograde halo is close to isotropic ($\sigma_R \approx \sigma_\phi \approx \sigma_z$), although $\sigma_R$ is always higher ($\sim100-110$~km/s) than $\sigma_\phi$ ($\sim80-90$~km/s), while $\sigma_z$ is the lowest ($\sim70$~km/s). We will further discuss the possible nature of the different halo components in Section~\ref{sec::compare}. 
Note that also the higher component GMMs for the VMP range do not contain rotation-supported disc components (see Fig.~\ref{fig:appendix}). Interestingly, the three-component model identified a component shown in the dashed aqua-coloured ellipse in the left panel of Fig.~\ref{fig:Best_fit_GMM}. This overdensity/asymmetry can also be seen in the top left panel of Fig.~\ref{fig:vr_vphi_space}, and is responsible for the (unexpected) tilt in the prograde halo ellipse in the VMP bin. We discuss possible origins for this population in Section~\ref{sec::vmpsubstructure}. 

\begin{figure*}
    \centering
    \includegraphics[width = 0.8\textwidth]{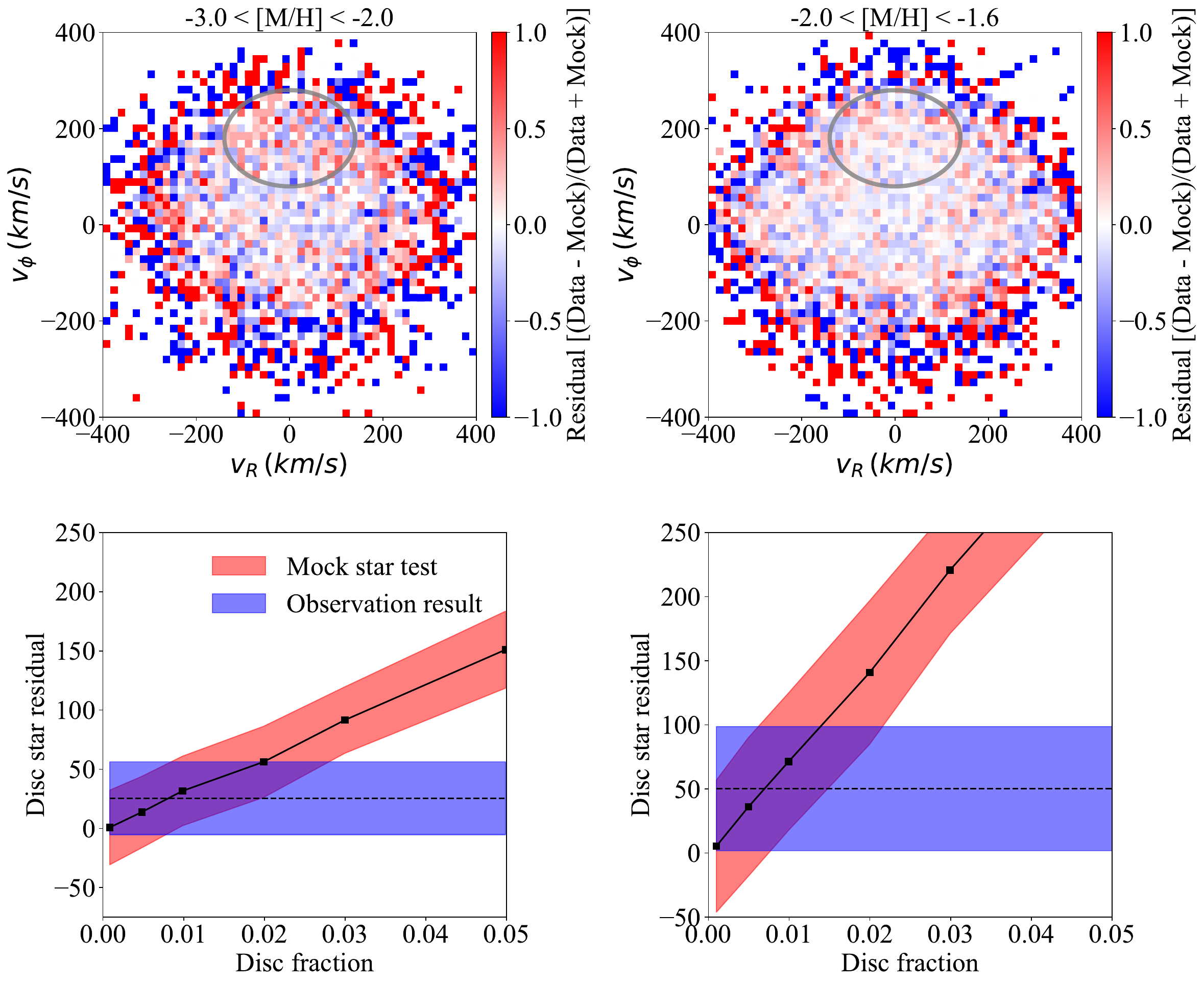}
    \caption{Top row gives the distribution of normalised residuals between the GMM and the data in the VMP (left) and IMP (right) bins. The normalised residual is defined as the number of mock stars subtracted from the number of observed stars divided by the sum of the observed and mock stars in each cell. The lack of strong systematic patterns, the low-amplitude and the small scatter of the residuals in the central regions verifies the validity of the GMM fit. The scatter in the periphery is larger due to random error. The grey ellipse indicates the region where stars with disc kinematics are expected in the $v_R-v_\phi$ plane. We count the residuals within the grey ellipse and show it as the dashed line in each of the lower panels. The lower panels illustrate the expected disc residual as a function of the disc fraction in the samples obtained using mock star tests. In the bottom row, the blue-shaded region denotes the $1\sigma$ uncertainty of the disc residual count computed using the Monte Carlo method. The solid black line and the red-shaded region show the residual and uncertainty of the mock sample (with manually added disc stars).}
    \label{fig:residual}
\end{figure*}

The GMM finds the same two halo components in the IMP bin, although the prograde halo now has more than 2.5x as many stars as the stationary halo, plus two additional sub-populations dominated by radial motions (the orange and aqua ellipses). These have very similar kinematics in $v_\phi$ and $v_z$, but are opposite in $v_R$, and are also found in the MP1 and MP2 bins. This kinematic signature suggests that they are likely connected to GS/E, the debris of a dwarf galaxy accreted by the MW in the last significant merger \citep{Belokurov_2018, Helmi_2018}. The debris of relatively high-mass accretion events is shown to be radialized efficiently over time, strongly increasing the eccentricities of the stellar orbits \citep[e.g.][]{Amorisco2017,Vasiliev2022}. Given the high resulting eccentricity of the bulk of the tidal debris, the pericentres and the apocentres of the GS/E debris are typically outside of the Solar neighbourhood, meaning that most of the GS/E the stars pass near the Sun with high radial velocity towards or away from the Galactic centre, causing two separate blobs in $v_R$. The bi-modal structure of the $v_R$ distribution of the GS/E debris had been anticipated \citep[see e.g. Fig 3 of][]{Deason2013}, can be clearly seen in the GMM residuals of \citet{Belokurov_2018} and is taken into account in the most rigorous models of the GS/E kinematics \citep[][]{Necib2019,Lancaster2019,Iorio2021}. The metallicities of our highly radial components are also in the expected range for GS/E -- for example, \citet{Myeong_2022} report mean and the dispersion of the GS/E metallicity distribution to be $\mu = -1.38$ and $\sigma = 0.20$ with a tail towards lower metallicities \citep[for other studies of the GS/E metalllcity distribution, see e.g.][]{Deason_2018,Feuillet2020,Naidu2020,Naidu2021}. Thus, we tag these two sub-populations shown in the orange and aqua ellipse as GS/E(1) (moving outward) and GS/E(2) (moving inward).

Note that given the estimated time of the GS/E accretion event of order of 8-11 Gyr ago \citep[see e.g.][]{Gallart_2019,Di_Matteo_2019,Belokurov_2020, Borre2022}, its stellar debris is expected to be phase-mixed and thus the fractional contribution of the positive and negative $v_R$ humps to be approximately the same. This appears to be the case for the two MP bins but not for the IMP bin, where the radial velocity distribution is clearly asymmetric: the negative $v_R$ component of the GS/E (aqua) appears to contain almost twice the number of stars compared to the positive one (orange). It is striking that this is in the same location as the third potential component identified in the VMP range. We further discuss this asymmetry in Section~\ref{sec::vmpsubstructure}.

\subsection{GMM residuals and the possible disc star fraction for VMP and IMP stars}\label{subsec::disc_fraction}

One of the limitations of GMM is that it does not guarantee to find all subpopulations in the sample, especially when the contribution is minor. Hence, we study the residuals between the GMM models and the data in the VMP and IMP regimes to examine whether there might be a disc population hiding that was not strong enough to be picked up by the GMM, but can be identified by a systematic pattern in the residuals. Our approach is as follows. We generate the same number of mock stars as the number of observed stars we have in each metallicity bin, where for each star, we generate $v_R$, $v_\phi$, and $v_z$ from the Gaussian distributions according to the parameters of the best-fit GMM model. Because we have used the Extreme Deconvolution algorithm for GMM fitting, the mock-generated stars are error-free. Therefore, we can not directly compare the generated data (without error) to the observed data (with error) and we need to assign measurement uncertainties to the generated stars to avoid a biased comparison. Matching the closest observed star for each mock star in the ($v_R,v_\phi,v_z$) space, we assign the measurement uncertainties of that observed star to the mock star, and then add random scatter to the generated velocities according to the assigned uncertainties.

We calculate the residual of the GMM fitting by subtracting the number count of mock stars from the observed count in each cell in the $(v_R,v_\phi,v_z)$ space. Fig.~\ref{fig:residual} shows the normalised residual distribution (top row) for the VMP and IMP bins (left and right, respectively). We model the thick disc by a 3D Gaussian distribution with mean $(v_R, v_\phi, v_z) = (0, 180, 0)$~km/s and a diagonal covariance matrix with dispersions $(\sigma_{R}, \sigma_{\phi}, \sigma_{z}) = (70, 50, 60)$~km/s, as highlighted by the grey ellipse representing $2\sigma$ (parameters inferred from the thick disc component in the MP2 bin, see Table~\ref{table:GMM_params}). By eye, there is no clear distinct residual pattern within the grey ellipse compared to the rest of the residuals. To quantify this, we count the residual between the mock and observed stars inside the $2\sigma$ region of the thick disc. We repeat the mock generation procedure 200 times to find the mean and uncertainties of the disc region residual. The residual count is $26\pm31$ and $50\pm48$ for VMP and IMP bin, respectively. This is represented as the dashed horizontal line in the lower panels of Fig~\ref{fig:residual}, where the blue band is the $1\sigma$ uncertainty region.

To understand the implication of these mean residuals, we compare them to values of residuals obtained in the presence of a mock disc population added to the data. We generate the same number of stars from the best-fit GMM model as we have data in the VMP and IMP bins. Then we add a disc population with varying size to that, following the velocity Gaussian of the thick disc as defined above. Similar to the previous procedure, we assign the measurement uncertainty to each star and add scatter in accordance with the uncertainty to mimic the observations. Repeating the previous steps for deriving the disc residual between the observation and the GMM model, we now calculate the disc residual between the mock sample (including the mock disc population) and the GMM model. By adjusting the number of disc stars inserted, we map out the disc residual as a function of the disc star fraction, shown by the solid line in Fig.~\ref{fig:residual}. The red band is the $1\sigma$ uncertainty (computed as before using the Monte Carlo method). By comparing the observed disc residual and the mock disc tests, we conclude that the fraction of the disc population in the VMP regime of our sample must be in the range of $0$ to $3\%$ (and $0$ to $2\%$ in the IMP range). Note that this fraction does not reflect the genuine fractional contribution of the disc population of all the Milky Way's VMP and IMP stars, but only the disc fraction in our sample suffering from the selection function introduced by the Gaia survey and the additional cuts we applied.

\subsection{Frozen GMM components}

\begin{figure}
    \centering
    \includegraphics[width = \columnwidth]{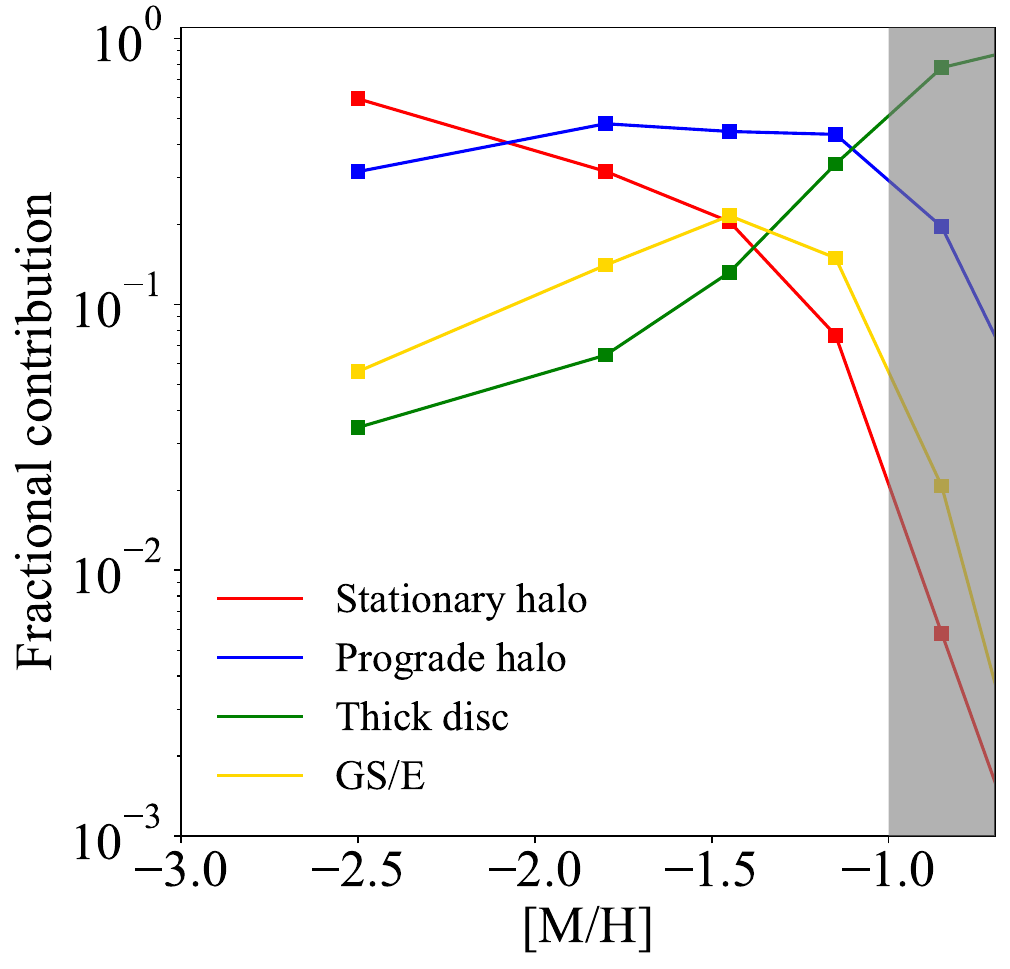}
    \caption{Fractional contribution of individual GMM components as a function of metallicity. The fractions are computed by freezing the mean and the covariance of each Gaussian in all metallicity bins and optimising for the weights only. We fix the GMM components according to the right panel in Fig.~\ref{fig:Best_fit_GMM}. The contributions from two GS/E components are added together. The metal-rich region  ($\mathrm{[M/H]}>-1$) is shaded as we focus the discussion on lower metallicities $\mathrm{[M/H]}\leq-1$. Note that the thick disc contribution dips quickly below $\sim10\%$ around [M/H]$\sim-1.5$ and remains low. The fractional contribution of the prograde halo is constant for $-2<$[M/H]$<-1.3$ but is reduced at [M/H]$<-2$ where the stationary halo's weight starts to dominate.}
    \label{fig:frozen}
\end{figure}

Here we describe an additional test of the GMM, which involves freezing the Gaussian components and only optimising the weights of each sub-population. We fit the same data (so again removing stars with $|z|>2.5$ kpc) for consistency with the previous analysis. We use the components of the GMM fitting for $-1.3<\mathrm{[M/H]}<-1.0$ in Fig.~\ref{fig:Best_fit_GMM} as the reference, because the five components recognised all have physical meaning and are in agreement with results in the literature \citep{Helmi_2020, Belokurov_2018, Necib2019, Lancaster2019, Iorio2021}. We modify the components slightly by setting the non-diagonal elements of the covariance matrix to zero for all components, as well as the mean $v_R$ and $v_{z}$ to zero except for the two GS/E components' mean $v_R$. Having fixed the mean and covariance of these five (corrected) components for all metallicity bins, we fit the models to the data to determine the weights associated with each component. 

The evolution of the weights of each component as a function of metallicity is shown in Fig.~\ref{fig:frozen}, where the contribution from the two GS/E components is added together. As expected, the two halo components dominate the metal-poor end. The stationary halo is the most significant component for [M/H]~$<-2.0$ and decreases in significance with increasing metallicity, while the prograde halo is the most dominant component for $-2.0 < \mathrm{[M/H]} < -1.0$, at a constant level of fractional contribution. The GS/E contribution peaks around [M/H]~$=-1.5$. The disc population is sub-dominant at low metallicity but appears around $\mathrm{[M/H]}\sim-1.6$ and becomes the main population by a factor of a few above [M/H]~$>-1.0$. 
The GMM assigns $3.4\%$ and $6.4\%$ of the total number of stars to the thick disc in the  VMP and IMP regimes, respectively. This fractional contribution of the thick disc is somewhat higher than indicated by the residual analysis in the previous sub-section, likely because the prograde halo in the VMP and IMP bins has evolved and has different properties in the MP2 bin. However, in this test with a fixed disc component in the GMM, the fractional contribution from the disc population is still very minor (even though we limit ourselves to the population relatively close to the plane, with $|z|<2.5$~kpc). 

Again it is worth noting that we do not claim that these fractional contributions are representative of the entire Milky Way -- they are biased by our selection function. However they are a good representation of our local neighbourhood, especially in comparison with other surveys that are also avoiding most of the plane of the Milky Way.

\section{Discussion}\label{sec::discussion}

\subsection{Impact of selection function}

We have applied several quality cuts thus causing a selection bias on top of the Gaia XP selection function. We have removed stars with low galactic latitude and high $\mathrm{E(B-V)}$ value. However, the disc is more prominent at low galactic latitude and therefore we are missing many of its stars. A similar effect occurs because we use giant stars -- giants are brighter than turn-off/dwarf stars and are therefore further away in a given magnitude range. On the other hand, we require stars to have a small fractional parallax uncertainty $\mathrm{fpu}<0.1$ to improve the quality of the kinematic measurements. This biases the sample closer to the Solar neighbourhood, which would in turn favour the disc population. We also remove stars with $|z|>2.5$~kpc when we do the GMM fitting, to focus more on stars close to the Galactic disc plane. The resulting sample after all cuts still overlaps significantly with the region expected to contain thick disc stars, see Fig.~\ref{fig:Sample_info} and Section~\ref{subsec::build_sample}.

Due to the selection effects described above, the fractional contributions of each Galactic component and of the disc in particular as computed here are only applicable to our sample and cannot immediately be generalised to the rest of the Galaxy. The full reconstruction of the selection function is beyond the scope of this work. Nevertheless, we believe that the selection biases caused by the strict quality cuts should not affect the arguments discussed above regarding the existence of a thick disc population in the VMP regime.
To verify this, we investigate the $v_R$-$v_\phi$ plane for sub-samples of stars in all the metallicity bins selected to have the same R and $|z|$ distributions as for the VMP sample (which is described with more detail in Appendix~\ref{Appendix::B}). Although the more metal-rich samples are now less confined to the disc plane, the transition from dispersion-dominated at low [M/H] to disc-dominated at higher [M/H] is still clearly visible (shown in Fig.~\ref{fig:appendixB}). Therefore, if there was a significant disc component in the VMP regime, we would have been able to identify it. 

It is clear that the sample employed in this work is not best suited to probe the existence of a VMP \textit{thin} disc, as most of the spatial region of the thin disc is excluded from our footprint. Such a population may be better studied with main-sequence stars instead, as faint but numerous dwarfs better sample the nearby low Galactic height regions. For example, 8 among the 11 UMP planar stars in \cite{Sestito_2019} are dwarf/turn-off stars. 
However, the absence of a VMP thick disc could constrain the origin of the VMP thin disc if it exists. One hypothetical origin of the disc-like VMP stars is that the Galactic disc formed at a very early epoch before the last significant merger \citep{Di_Matteo_2020}. Necessarily, the early thin disc would experience dynamical heating, and the stars therein would be splashed into hotter orbits (e.g., thick disc and halo). This is discussed in the next section in more detail. Hence, the likelihood of finding a VMP thin disc that formed before the thick disc and survived appears small.

\subsection{Limitations of the GMM}

Gaussian Mixture Modelling attempts to represent each sub-population in the sample as a Gaussian distribution in the feature space. This decomposition is not physically motivated and is clearly an over-simplification of the actual distribution function of the components. For example, the thin disc velocity distribution is strongly non-Gaussian due to e.g. the effects of the asymptotic drift. Accordingly, we have only attempted unrestricted GMM fitting in the range of $-3<\mathrm{[M/H]}<-1$, where the thin disc population is insignificant. At higher metallicities, if set free, the GMM algorithm would waste many components to describe the thin disc behaviour. While the thin disc is the clearest example of a non-Gaussian behaviour, other components are also not guaranteed to be well-modelled by a single Gaussian. For example, the GS/E tidal debris appears bi-modal in the $v_R$ dimension across a wide range of Galactocentric distances, although each radial velocity hump is approximately a Gaussian \citep[][]{Necib2019,Lancaster2019,Iorio2021}. Additionally, a contribution from the stars trapped in resonances with the bar would also make the velocity distributions asymmetric and non-Gaussian \citep[see e.g.][]{Dillamore_2023bar}.

To overcome this, instead of the GMM, we could fit a mixture model with dynamical distribution functions that describes the kinematics of the disc and halo stars more accurately \citep[e.g. quasi-isothermal distribution function for a disc][]{Binney_2010}. With such kinematic/dynamical modelling, we could also extend the sample to stars without radial velocity measurements. Distribution function approach naturally allows to marginalise over the missing radial velocity but it would be more computationally expensive, and is beyond the scope of the current work. Another advantage of using physically-motivated dynamical models would be that, as opposed to the GMM, the results are guaranteed to be interpretable. 

As Fig.~\ref{fig:Best_fit_GMM} demonstrates, the best-fit model requires the prograde halo component to have a tilt in the VMP regime. This unphysical result could be caused by a dynamical substructure in the Milky Way that resides in a region of $v_R$-$v_{\phi}$ space similar to GS/E(2). This could also provide an explanation for the asymmetry of the GS/E(1) and GS/E(2) weights in the IMP regime (we will discuss this structure further in Section~\ref{sec::vmpsubstructure}).  It is unclear whether such tilt can be reproduced within the dynamical modelling framework, under the assumptions of equilibrium and axi-symmetry. Note that these assumptions have been demonstrated not to hold exactly in the MW today: the Galaxy is currently out of equilibrium near the Sun \citep[see e.g.][]{Antoja2018}, and further in the halo  \citep[due to an interaction with the Large Magellanic Cloud, see][]{Erkal_2019}. 

\begin{figure*}
    \centering
    \includegraphics[width = 0.9\textwidth]{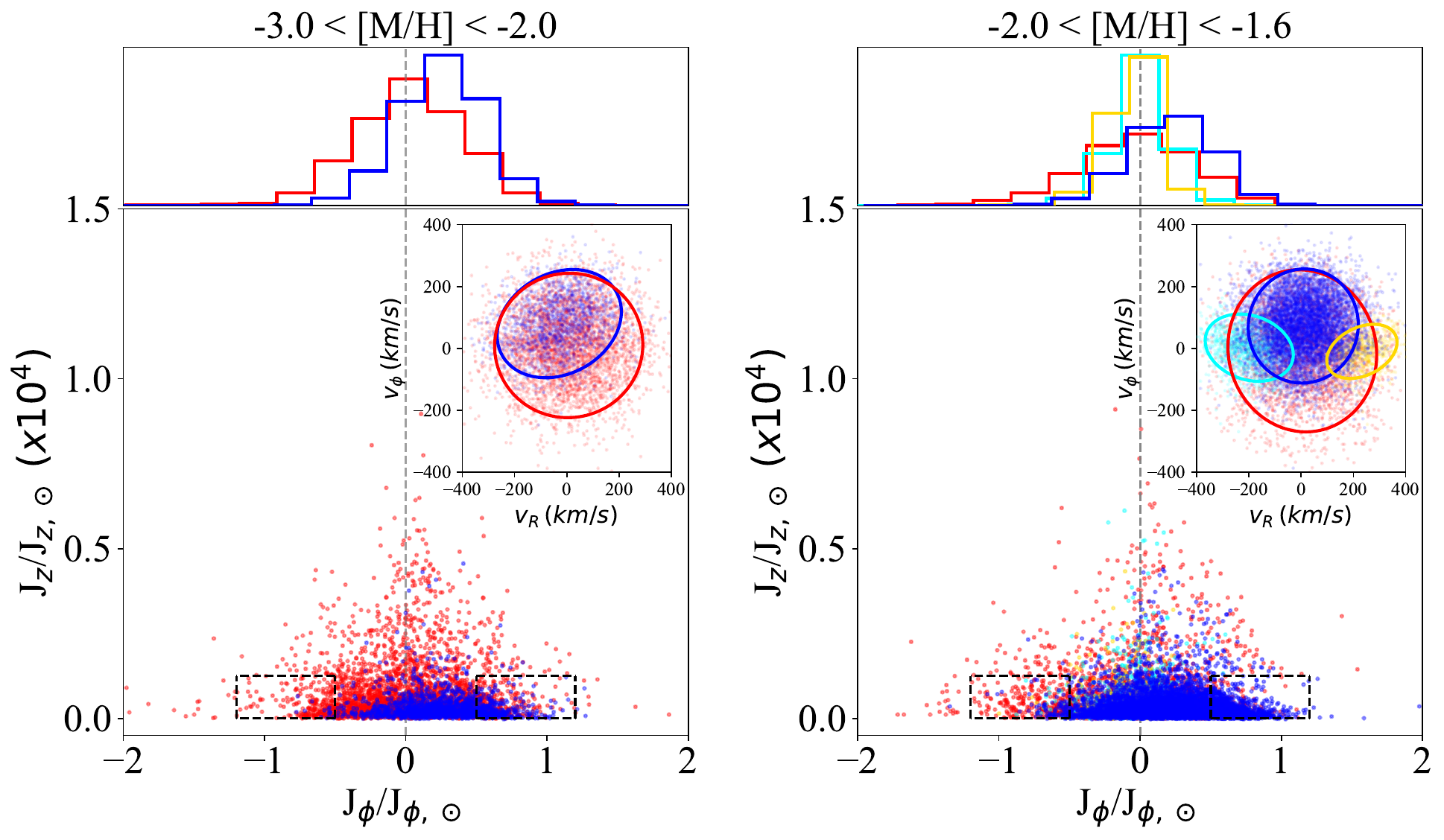}
    \caption{Distribution of low-metallicity stars in the  $J_\phi - J_z$ action space where $J_{\phi,\odot}$, and $J_{z,\odot}$ are the azimuthal and the vertical actions of the Sun. This can be compared directly to the results presented in \citet{Sestito_2019, Sestito_2020}. We assign the membership for each star in the VMP (left column) and IMP (right column) bin by generating mock stars using the procedure described in Section~\ref{subsec::disc_fraction} and finding the closest match in the $v_R-v_{\phi}-v_z$ space.  We then assign the membership of the closest matched mock star to the observed star. The membership classification is illustrated in the $v_R-v_{\phi}$ inserts  The stars are projected into the $J_\phi - J_z$ space and coloured by the membership assigned. The colour-coding is the same as in Fig.~\ref{fig:Best_fit_GMM}. Two black-dashed boxes represent the prograde and the retrograde disc-like orbits as proposed by \citet{Sestito_2019, Sestito_2020}. The histograms on the top of each panel is the distribution of $J_\phi$ for sub-populations (following the same colour notation). A grey-dashed line is drawn on $J_\phi=0$ to emphasise the overdensity in the prograde population.}
    \label{fig:action_action_space}
\end{figure*}

The quality of the GMM fitting can also be affected by substructures in the Milky Way (e.g. globular clusters, dwarf galaxies, stellar streams, shells and chevrons). The known satellites are easy to deal with -- we removed stars within $1\,\mathrm{deg^2}$ of the on-sky locations of all known Galactic globular clusters and dwarf galaxies. Due to strict cuts on the fractional parallax uncertainty and extinction, the sample analysed here is limited to a relatively small volume around the Sun. This mitigates the effects of unmixed tidal debris in our analysis. Although many local halo substructures have been reported in the literature \citep[e.g.][]{Koppelman_2018,Myeong_2018a,Yuan2018}, their fractional contribution may be rather small \citep[see][]{Naidu2020,Myeong_2022}.

Another caveat of GMM is that it is often not efficient in identifying under-represented populations. As is stated throughout the paper, given the data sample in hand, we can not rule out the presence of a minor disc-like component. Instead, we address the issue of the GMM being mostly sensitive to components with large fractional contributions by analysing the residual of the GMM fitting. Further, we extract the disc contribution in the IMP and VMP bins by freezing a disc-like population when running the GMM fitting, and the small weighting factor associated with the disc-like population reinforces that the \textit{thick} disc stars have a minor contribution in the VMP and IMP regime.

\subsection{Comparison to previously reported observations}\label{sec::compare}

\cite{Sestito_2019} analysed the kinematics and dynamics of all ultra metal-poor (UMP, [Fe/H]~$<-4.0$) stars available at the time. They found that 11 out of 42 ($\sim26\%$) UMP stars are confined within $3\,\mathrm{kpc}$ of the Milky Way plane throughout their orbital lifetime. Furthermore, 10 of these 11 UMP stars are in prograde orbits ($v_\phi > 0$). \cite{Sestito_2020} reach a similar conclusion using a much larger collection of 1027 VMP and extremely metal-poor (EMP, [Fe/H]~$<-3.0$) stars in the sample combining Pristine survey spectroscopic follow-up data \citep{Starkenburg_2017_Pristine, Aguado_2019} and LAMOST spectroscopy \citep{Deng_2012_LAMOST, Li_2018}. They show that $\approx31\%$ of stars with $\mathrm{|z|}<3$~kpc observed today never travel outside of $3\,\mathrm{kpc}$ of the disc plane. They also study their sample in the $J_\phi-J_z$ projection of the action space, where $J_\phi$ is the azimuthal action (or angular momentum, $L_z$) and $J_z$ is the vertical action. \cite{Sestito_2020} find that the number of stars in prograde disc-like orbits, i.e. high $J_\phi$ and low $J_z$, is $5\sigma$ greater than that of the stars in retrograde disc-like orbits. 

We perform an orbital analysis similar to that of \cite{Sestito_2019,Sestito_2020} and study the behaviour of our VMP and IMP stars in the action space, as illustrated in Fig.~\ref{fig:action_action_space}. We find the fraction of VMP and IMP stars with the present day $\mathrm{|z|}<3$~kpc and $z_{\rm max}$ $\leq$ $3\,\mathrm{kpc}$ are $22\%$ and $28\%$ respectively, not too different from \cite{Sestito_2020}. In the $J_\phi-J_z$ plane, we count stars in the black dashed boxes that represent disc-like prograde and retrograde orbits. The region corresponding to low-$z_{\rm max}$ corotating stars is $13.6\sigma$ (VMP) and $28\sigma$ (IMP) over-dense compared to its retrograde counterpart. We conclude that for stars with $-3<$~[M/H]~$<-1.6$ in our sample, the prograde/retrograde asymmetry is much stronger compared to that found by \cite{Sestito_2019,Sestito_2020}. Note however that the footprints and the selection functions on these studies are very different (e.g. our sample is all-sky while their samples are mostly limited to the Northern hemisphere and avoid the disc regions). 

Interestingly, \citet{Dillamore_2023bar} show that the prograde/retrograde asymmetry and an over-density in $J_\phi-J_z$ space similar to that observed here and in \cite{Sestito_2019,Sestito_2020} can arise naturally in the extended Solar neighbourhood in the presence of a rotating bar. Further exploration is needed to quantify exactly how much of the asymmetry is due to the bar. 

In the action space in Fig.~\ref{fig:action_action_space} we colour-code stars by their membership in the detected GMM components. Unsurprisingly, this shows that stars belonging to the prograde halo component (blue) are the main cause of the asymmetry between the prograde and the retrograde obits near the plane. Thus, in our study, rather than coming from an intact, rotation supported disc, the asymmetry can be explained by the contribution of a kinematically hot population with a small net rotation. \citet{Belokurov_2022} see a very similar behaviour in the in-situ population they call Aurora. In the APOGEE  DR17 data, Aurora stars start to dominate the in-situ component below [Fe/H]~$=-1$. This is in a good agreement with our results, see for example Fig.~\ref{fig:frozen} where the fractional contribution of the prograde halo is the largest of the four components for $-2.0 \lesssim $~[M/H]~$\lesssim-1.0$. Without a detailed chemical information it is impossible to be certain that the prograde halo identified here is the Aurora of \citet{Belokurov_2022}, even at [M/H]~$>-1.6$. In fact it seems likely that the prograde halo Gaussian component would absorb some of the stars classified by \citet{Belokurov_2022} as accreted. Nonetheless, our analysis indicates that a component similar to Aurora can be detected in the {\it Gaia} XP+RVS data and that its contribution remains relatively high at metallicities lower than studied before, i.e. $-2.5<$~[M/H]$<-1.5$. 

We therefore tentatively associate our prograde halo population with the pre-disc in-situ population of the MW. For [M/H]~$<-2.0$, the stationary halo component becomes dominant in our GMM (Fig.~\ref{fig:frozen}). Given its low metallicity, high velocity dispersion (higher than that of the prograde component) and negligible rotation, we suggest that this component might be dominated by accreted stars from a large number of small accretion events. \citet{Arentsen_2024} argue along similar lines for the origin of metal-poor stars in the inner few kpc of the MW from the Pristine Inner Galaxy Survey \citep{Arentsen_PIGS_II_2020}. They find that the rotational velocity decreases as function of metallicity, and interpret this as a transition from in-situ to accretion-dominated. 

\citet{Bellazzini_2024} used a sample of giant stars with photometric metallicities derived from Gaia DR3 synthetic Str\"omgren photometry to investigate the dynamics of low-metallicity stars. By inspecting the angular momentum distribution of prograde and retrograde stars for metallicities below $-1.5$, they also found a prevalence of prograde over retrograde stars and interpret this as evidence of the presence of a low metallicity disc (or the ``seed'' of the disc). We use a very similar dataset in this work with very similar quantitative results, but have a rather different interpretation of what these numbers mean. We argued above that this asymmetry can be the result of i) a prograde halo component (which in our analysis has v$/\sigma \sim 1$, which would not usually be classified as a disc) and/or ii) bar-driven resonances.

\cite{Di_Matteo_2020} compile a sample of 54 EMP and VMP stars and add UMP stars from \cite{Sestito_2019}, metal-poor and metal-rich stars from \cite{Nissen_Schuster_2010} and APOGEE stars, and argue that the disc population exists across a wide metallicity range $-6<\mathrm{[M/H]}<1$. \cite{Di_Matteo_2020} show that these stars all occupy a particular region in the Toomre diagram, i.e. where the orbital motion is dominated by prograde rotation.
However, the small sample size and the unknown selection effects of \cite{Di_Matteo_2020} make it difficult to interpret their findings. In our analysis, with much larger sample size and a consistent selection function across VMP to the metal-rich regime, the detailed behaviour of the stellar kinematics in various metallicity bins appears to be significantly different to that presented in \citet{Di_Matteo_2020}. More specifically, we find little observational evidence in support of their hypothesis of the disc population being ubiquitous at all metallicities. 

\citet{Nepal_2024} analysed the chrono-chemo-kinematic evolution of the Milky Way using the parameters derived from \textit{Gaia} XP+RVS data and found a disc population older than $13$~Gyr, but in metal-rich stars rather than metal-poor stars. Therefore, the result does not contradict our conclusion directly. There are various origins and interpretations of the metal-rich old disc population, such as an early formed disc and/or stars migrated from the inner Galaxy due to bar activities \citep[e.g. ][]{Li_2023}. Note however that effects associated with the age uncertainty could also cause produce a signal similar to that presented in \citet{Nepal_2024}. We discuss the formation time of the Milky Way's disc in more detail in Zhang et al. (in prep) by exploring the chrono-kinematic signatures of the inner Galaxy.

The proposed metal-weak thick disc (MWTD) consists of stars with thick disc kinematics but with lower metallicities  \citep{Morrison_1990, Norris_1985}. Recent views have evolved, and many works have argued that the MWTD has kinematic and chemical properties distinct from the canonical thick disk \citep{Carollo_2010, Carollo_2019, AnBeers_2020, Mardini_2022}. The metallicity range claimed for the MWTD in these works is consistent with that probed by our analysis. However, as we do not have $\mathrm{[\alpha/Fe]}$ measurements in our sample, and due to our limited spatial coverage, we cannot unambiguously confirm that the thick disc component we find in the MP1 and MP2 bins is the same as the MWTD in \cite{Carollo_2019} and \cite{AnBeers_2020}. Also, the presence of a prograde halo complicates the picture of the Milky Way in the metal-poor regime. Therefore, the conventional component membership assignment method based only on the kinematic property of a thin disc, thick disc, and stationary halo (e.g. \citealt{Mardini_2022, Li_Zhao_2017}) might need revision. The prograde halo, i.e. a kinematically hot component with a small net spin, can account for an excess of stars with positive angular momentum assigned to the disc by the above studies.

\subsection{Circular-like orbits in halo populations}\label{sec::circ}

\begin{figure}
    \centering
    \includegraphics[width = \columnwidth]{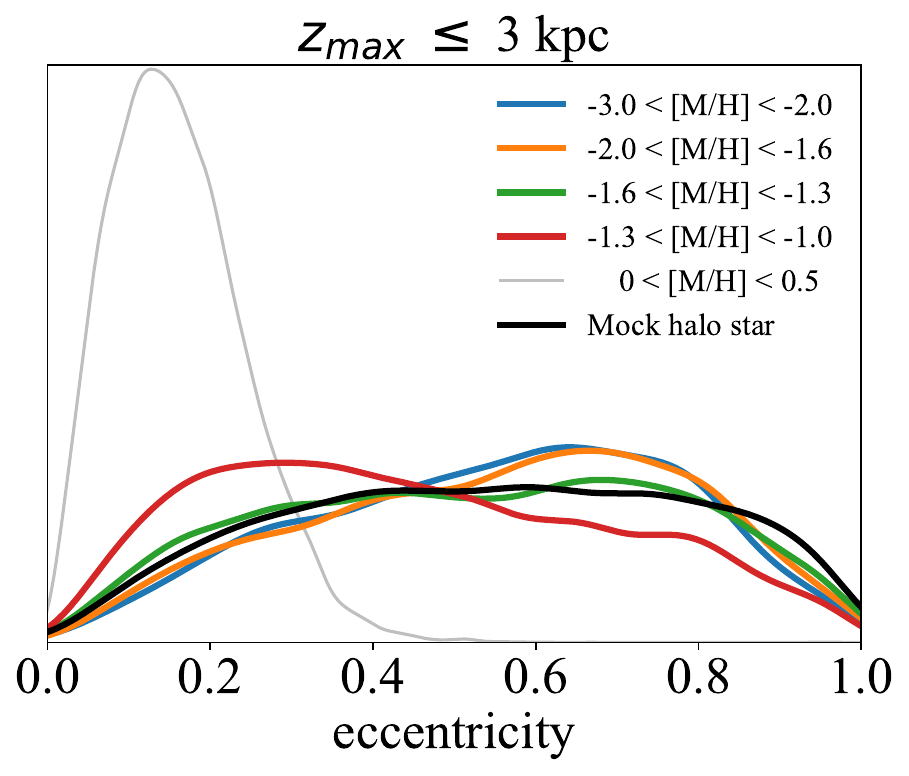}
    \includegraphics[width = \columnwidth]{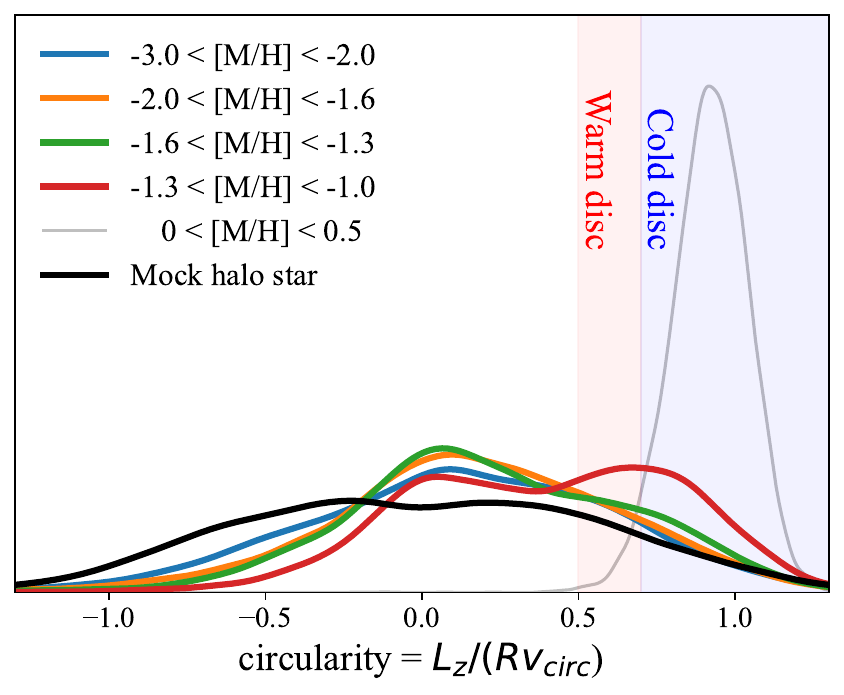}
    \caption{The eccentricity (top panel) and circularity (bottom panel) distributions of stars with $-3<\mathrm{[M/H]}<-1$ in different metallicity bins (blue, orange, green, and red lines in order of increasing [M/H]). Also shown are particles from a mock non-rotating halo population (thick black line). The mock particles are drawn from the isotropic NFW distribution function. The eccentricity distribution of metal-rich stars ($0<\mathrm{[M/H]}<0.5$) is also plotted in grey as a reference for the behaviour of a disc population. All distributions are computed using  Kernel Density Estimation. The blue and red-shaded regions in the bottom panel are the cold and warm disc selection criteria defined in \citet{Sotillo-Ramos_2023}.}
    \label{fig:e-distribution}
\end{figure}

Prograde motion is a necessary but not sufficient condition to show the existence of a disc. Accordingly, we look for additional evidence of a VMP disc-like population in other orbital parameters, namely in distributions of eccentricity $e$ and maximal vertical excursion $z_{\mathrm{max}}$, which have been used in the literature, as well as the orbital circularity.

\subsubsection{Eccentricity and $z_{\mathrm{max}}$}

Fig.~\ref{fig:e-distribution} shows eccentricity distributions for stars with $z_{\mathrm{max}}\leq 3$~kpc in different metallicity bins. The entire orbits of these stars remain relatively close to the galactic plane, therefore, they are a perfect candidate for the stellar disc population. However, as shown in the Figure, these distributions are lopsided towards high eccentricity, $e>0.5$ in all metal-poor bins. Only the most metal-rich bin considered, $-1.3<\mathrm{[M/H]}<-1$ (red), exhibits the prevalence of orbits with $e<0.5$. Combined with the qualitative arguments above, this further supports the earlier claims that in the Milky Way, $\mathrm{[M/H]}\sim-1.3$ is the lowest metallicity where an intact, rotation-supported disc can be detected. 

What is expected for the eccentricity distribution of a pure halo population? For comparison, we generate a sample of mock halo stars using an isotropic Navarro–Frenk–White (NFW, \citealt{NFW_profile}) distribution function \citep{Widrow_2000}. The NFW potential we adopt for this exercise has a scale radius of $16$~kpc \citep{Bovy_2015} and a mass normalisation that supports the circular velocity at solar radius, $v_{\mathrm{circ}}(R_{\odot})\sim220$~km/s. By design, this mock population has no net rotation. As our investigation focuses on the solar neighbourhood where the disc potential is crucial, we evolve these mock halos stars in a specially manufactured potential that mimics the real environment while avoiding the non-adiabatic transition of the distribution function. We set a time-evolving potential that only consists of the same NFW potential at time $\mathrm{t}=0$ (beginning of the orbit integration). We define a disc potential with the same parameters (except for the mass normalisation) as in \cite{Bovy_2015}. The disc starts to grow from $\mathrm{t}=0$ while the strength of the NFW potential decreases so that the mass normalisation roughly preserves $v_{\mathrm{circ}}(R_{\odot})\sim220$~km/s. The potential becomes time-independent after $\mathrm{t}=1$~Gyr, and the final state of the potential is constituted by the same halo and disc as in \cite{Bovy_2015}. 

We use the orbit integration routine with an adaptive-step size integrator in \texttt{AGAMA} \citep{Vasiliev_2019} to evolve the mock-generated halo population in the potential described above, up to $8$~Gyr. Visually inspecting the distribution function at different time slices, we find that the population reaches a new equilibrium before $\mathrm{t}=2$~Gyr, and therefore, we compute the orbital eccentricity using orbital trajectories between $2<\mathrm{t\,(Gyr)}<8$. We assume the final time snapshot of the generated halo stars as the moment of observation. To ensure a fair comparison between the mock sample and the real sample, we match the $R-z$ distribution of the mock stars to that of the real stars with metallicity between $-3<$~[M/H]~$<-1$ (see their $R-z$ distribution in the mid-panel of Fig.~\ref{fig:Sample_info}). For every real star in the sample, we find a mock star that has the closest match to the real star in the $R-z$ plane, and we discard all the mock stars that fail to become the closest match to any of the real stars. As a result, the $R-z$ distribution of the mock stars matches the observed distribution in the {\it Gaia} XP+RVS sample. Finally, we also apply the $z_{\mathrm{max}}<3$~kpc cut to the mock halo sample to isolate the ``disc star candidates''.

The eccentricity distribution of this halo mock sample is shown as the solid black line in Fig.~\ref{fig:e-distribution}. It matches the distributions of observed metal-poor stars ([M/H] $<-1.3$) well, including a similar fraction of stars with relatively low eccentricity orbits, discy ($e < 0.4$). The similarity between the eccentricity distributions of the mock halo particles and observed low-metallicity stars is striking, which demonstrates that the low eccentricity (and low $z_\mathrm{max}$) stars in our sample can naturally occur in a pure halo component without the need for adding a disc. 

\subsubsection{Circularity}

Several authors have used orbital circularity to assign stars to the disc(s). For example, \citet{Sotillo-Ramos_2023} use a combination of circularity and Galactocentric radius R to classify different Galactic components in MW-like galaxies found in the TNG50 cosmological simulation suite. They define the circularity as $L_z/(rv_{circ})$, with $v_{circ}$ the circular velocity at galactocentric radius $r$. They adopt $\mathrm{circularity}>0.7$ to assign stars to the cold disc, and assign stars with $0.5<\mathrm{circularity}<0.7$ and $3.5 < R < 6\times\mathrm{Disc\, Scale \,Length}$ (for the Milky Way that roughly corresponds to  $3.5 < R <13$~kpc) to the warm disc. Only a very small fraction of our sample falls outside this $R$ range (see Figure~\ref{fig:Sample_info}).
Under these definitions, and after applying a cut on heliocentric distance $<5.5$~kpc (a volume similar to our selection, although we cut out the disc plane), \citet{Sotillo-Ramos_2023} find that $\sim20$ per cent of VMP stars in the simulated MW analogues can be assigned to the disc.

To make a comparison with their predictions, we compute circularity for all stars in our sample as well as the mock halo sample (for convenience of comparison, we compute circularity using cylindrical galactocentric radius $R$). We present the circularity distribution for each metallicity bin and for the mock halo sample in the bottom panel of Fig.~\ref{fig:e-distribution}. The red and blue-shaded areas are the cold and warm disc regions, following the definition of \citet{Sotillo-Ramos_2023}. A non-negligible fraction of stars in all metallicity bins resides in the high-circularity regions. Quantitatively, $24$ per cent of VMP stars and $25$ per cent of IMP stars are assigned to the disc (cold and warm), which is similar to the fraction of VMP disc stars in the simulations of \citealt{Sotillo-Ramos_2023}. However, $21$ per cent of the mock halo population is assigned to the disc as well. Note that, by definition, there is no disc present in the latter sample.

The circularity distribution of the mock halo population deviates more from that of observed VMP and IMP stars than the eccentricity distribution did. This is because the circularity not only depends on the eccentricity but also takes into account the contribution from the orbital direction. The net prograde rotation in our VMP and IMP samples results in an asymmetry in the circularity distributions, whereas the mock halo population does not include any rotation and is as a result symmetric. Fewer stars will therefore be assigned to the disc component in the mock halo population. We conclude that using circularity alone appears to be insufficient to show the existence of a disc population, since a large fraction of halo stars would also be included in the selection, this fraction is higher still if the halo has a small net prograde rotation.

\subsection{Negative $\mathbf{v_R}$ substructure}\label{sec::vmpsubstructure}

Circled by the dashed-aqua ellipse in the left panel of Fig.~\ref{fig:Best_fit_GMM}, a clear overdensity of stars with high eccentricity and small net rotation exist in the very metal-poor regime. It looks kinematically similar to the inward-moving phase (negative $v_R$) of the GS/E, but weirdly, the outward phase is missing, meaning that this population of stars only moves towards the Galactic centre and never comes back. The IMP bin also has an asymmetry between negative $v_R$ and positive $v_R$ among the GS/E components, which might be connected to the negative $v_R$ overdensity in the VMP range. 

Hints of a $v_R$ asymmetry in the GS/E debris were already found by \citet{Belokurov_2018}, and in the analysis of phase-space chevrons among halo stars in \citet{Belokurov_wrinkles2023} and \citet{Donlon2023}, the chevron at negative $v_R$ also appears to be more prominent. This overdensity is only recognisable between $-3<\mathrm{[M/H]}<-1.6$ in Fig.~\ref{fig:Best_fit_GMM}, but it could also extend to more metal-rich ends. The overdensity becomes indistinguishable as GS/E starts to dominate the population in that region of the velocity space for more metal-rich bins. The possible causes of the overdensity are recent accretion event debris that is not fully phase-mixed, bar-resonances affecting halo stars, or the selection function (the latter is less likely). We will investigate these possible explanations in the future.

\section{Simulations of galaxy formation and the emergence of discs}\label{sec::simulation}

Our observational study is linked to the following two questions on the galaxies' salient transformation phases. How early in the life of a galaxy can a stellar disc form? How easily can it subsequently get destroyed? High-resolution hydro-dynamical numerical simulations of galaxy evolution can be interrogated to provide clarity on the emergence and the destruction of stellar discs in Milky Way-like galaxies, at least within the current setup of our structure formation paradigm.

The timing of the disc formation has been the focus of several simulations-based studies most recently.  \citet{Belokurov_2022} introduce the term {\it spin-up} to describe a systematic increase in the median rotational velocity of the MW stars as a function of metallicity. They show that the mode of the stellar azimuthal velocity distribution changes from values just above 0 km/s to $\sim150$ km/s between [Fe/H]~$\approx-1.3$ and [Fe/H]~$\approx-1$. They compare this APOGEE DR17-based measurement to the behaviour of the redshift $z=0$ median spin of stellar particles in the MW-like galaxies in the Auriga (4 systems) and FIRE (7 systems) simulations. All galaxies in both suites go through the spin-up phase, and, similarly to the stars in the MW, do it relatively fast, i.e. covering a little range of metallicities. However in simulations this transition happens at significantly higher metallicities compared to the MW observations. Analysing the spin-up lookback times, \citet{Belokurov_2022} show that the FIRE galaxies start to form stellar discs rather late, i.e. 6-9 Gyr ago. Auriga galaxies spin up earlier, 9-11 Gyr ago. \citet{Belokurov_2022} explore the state of the FIRE stellar distributions at redshifts preceding the spin-up and demonstrate that they are irregular and lumpy, characteristically non disc-like. Curiously, in all simulations considered, the stellar particles in place before the spin-up show small systematic prograde rotation at redshift $z=0$, similar to our {\it prograde halo} component.

\citet{McCluskey2023} use an extended set of 11 MW-like galaxies as part of the FIRE-2 simulations to conduct an in-depth study of the evolution of the stellar kinematics across all ages, from birth to redshift $z=0$. They detect no primordial discs -- the FIRE galaxies start without coherent rotation at lookback times greater than 10 Gyr. \citet{McCluskey2023} show that the mock MWs subsequently go through a relatively rapid disc emergence phase when the median azimuthal velocity increases together with $v_{\phi}/\sigma_z$, i.e. the ratio of rotation velocity to vertical velocity dispersion, meant to quantify the amount of rotational support (similar to the quantity plotted in Fig.~\ref{fig:V_sigma} of our study). Importantly, they demonstrate that the kinematic behaviour of the galaxy before, during and after the disc emergence remains the same viewed either at the time of formation or at present day. Note that according to \citet{McCluskey2023}, while the kinematics of the stellar particles younger than 10 Gyr remains largely unchanged, the oldest stars suffer the largest amount of heating. Moreover, the oldest stars born with zero rotation gain a small amount of spin by the present day (consistent with our measurement of the prograde halo).

\citet{Semenov_2023} look at the disc formation in the MW-mass galaxies in the Illustris TNG50 suite. While TNG50 has lower resolution compared to FIRE and Auriga, it is sufficient to resolve and study the kinematic history of MW-like galaxies. The obvious benefit of using TNG50 is that it offers a larger sample of systems to study, i.e. $\sim100$ compared to $\sim10-20$ in zoom-in suites. To overcome potential metallicity biases present in TNG50, \citet{Semenov_2023} re-calibrate the metallicity distributions of the mock MWs using the APOGEE observations reported in \citet{Belokurov_2022}. They report that the rapid spin-up phase is ubiquitous in TNG50. However, even after the re-calibration, \citet{Semenov_2023} find that the simulated galaxies spin up later compared to the observations: only 10\% of the systems considered have the spin-up metallicity consistent with the measurements of \citet{Belokurov_2022}, implying that the MW is not a typical galaxy for its total mass. While it is true that the bulk of the TNG50 MWs spin up at significantly higher metallicities, in many of these, the difference in the time of the disc emergence is small. This subtlety is explained \citet{Semenov_2023} and is attributed to a very rapid self-enrichment phase most MW-like galaxies experience at high redshift. \citet{Semenov_2023} discover the link between the disc spin-up time and the galaxy's accretion history: the mock MWs with an early spin-up at low metallicity are those that assemble the fastest. These authors also show that late spin-up galaxies suffer significant destructive mergers at late times, in addition to failing to form a prominent stellar disc early.

\citet{Dillamore_2023spin} explore the connection between the time a dominant stellar disc emerges and the mass assembly history of 18 galaxies in the zoomed-in hydrodynamical ARTEMIS suite \citep[][]{Font2020}. Their study focuses on approximately half of all ARTEMIS systems, a sub-set that have a disc at redshift $z=0$ and thus are closer analogues of the MW. \citet{Dillamore_2023spin} find a clear correlation between the spin-up time and the host's dark matter halo mass at the lookback time of 12 Gyr: the higher DM masses at early times imply faster disc emergence, Expressed differently, the ARTEMIS galaxies form a dominant stellar disc when their DM halo masses reach $\sim 6\times10^{11}M_{\odot}$. Of the 6 galaxies with the earliest spin-up times ($>$9 Gyr ago), five are objects containing a GS/E-like structure in their accreted stellar halo. This is in agreement with the findings by \citet{Fattahi2019} and \citet{Dillamore2022}, who show that the presence of the GS/E in the galaxy today conditions  the mass assembly history to peak early. \citet{Dillamore_2023spin} show that such an assembly history biases the galaxy to have a lower accreted stellar mass fraction as the number of massive mergers is reduced for a sample of galaxies with fixed redshift $z=0$ mass. This also explains why MW analogues -- selected to have GS/E-like events -- show comparatively lower DM spin.

Taken together, the studies of the currently available numerical simulation suites discussed above present a coherent picture of the disc emergence and survival in the MW-mass galaxies. Simulated galaxies do not possess fast-spinning, stable and dominant stellar discs at metallicities significantly below [Fe/H]~$\approx-1$. In other words, metallicity of [Fe/H]~$\approx-1$, corresponding to the spin-up metallicity of the MW, is very close to the lowest metallicity at which stable, dominant stellar discs are observed in simulations. At lower metallicities, i.e. [Fe/H]~$<-1$, the in-situ stellar particles predating the spin-up phase are born in a messy state without coherent rotation. DM halos hosting mock MWs with the lowest spin-up metallicities assemble their masses the earliest. Their accretion histories are also "tuned down" to preserve the primordial disc intact with only moderate heating. The MW did survive its most significant merger, that with the GS/E progenitor, even though its pre-existing disc got splashed \citep[][]{Gallart_2019,Di_Matteo_2019,Belokurov_2020}. The simulations reveal that splashing is not the only consequence of massive accretion events: the pre-existing discs can be also tilted \citep[][]{Dillamore2022,Chandra2023}.

\section{Conclusions}\label{sec::conclusions}

In this work, we combine precise phase-space measurements from \textit{Gaia'}s astrometry and spectroscopy with a large sample of homogeneous metallicities derived from the low-resolution \textit{Gaia} XP spectra \citep{Andrae_2023}, covering $-3<\mathrm{[M/H]}<0.5$, to investigate the presence of a rotation-supported ($v/\sigma>1$) structure in the VMP ([M/H]~$< -2.0$) regime based on the 3D velocity distributions in various metallicity ranges.

The $v_R$-$v_{\phi}$ space for stars in the extended Solar neighbourhood does not show a strong, stand-alone rotation-supported disc population in the VMP regime (see Fig.~\ref{fig:vr_vphi_space}). We approximate the stellar velocity distribution with a Gaussian mixture model (GMM) and find that it does not identify a distinct VMP disc component (see Fig.~\ref{fig:Best_fit_GMM}). Instead, we find tentative evidence of two halo populations. One is a prograde halo-like component with a net $v_\phi \sim 80$~km/s, and the other is a static halo with larger velocity dispersion. Note that our analysis does not exclude the existence of a sub-dominant disc-like population at low metallicities.

The earliest rotation-supported disc population is detected among stars between $-1.6<\mathrm{[M/H]}<-1.3$, which roughly agrees with the previous observational constraints on the metallicity scale at which the Galactic disc formed \citep[][]{Belokurov_2022,Chandra2023} as well as the numerical estimates \citep[see][]{Semenov_2023,Dillamore_2023spin}. We also identify Gaussian components connected to the GS/E merger \citep{Helmi_2018, Belokurov_2018} for [M/H]~$>-2.0$, e.g. two lobes with low $v_\phi$ and strongly positive or negative $v_R$ in agreement with previous detailed analyses of the GS/E kinematics \citep[see e.g.][]{Necib2019,Lancaster2019,Iorio2021}. 

Fixing the GMM components to the combination of a stationary halo, a prograde halo, a thick disc and the GS/E, we show the transition from a dispersion-dominated to a disc-dominated Galaxy around [M/H]$\approx-1.3$ (see Fig.~\ref{fig:frozen}). The prograde halo is the main component for $-2.0 <$~[M/H]~$<-1.0$, and we find that it has similar kinematic properties to \textit{Aurora}, the ancient in-situ halo of the Milky Way \citep{Belokurov_2022}. The stationary halo becomes the main component for [M/H]~$<-2.0$, which we tentatively associate with predominantly accreted stars. 

We used the residuals between the best-fit GMM components and the observed VMP and IMP samples to place limits on the maximum thick disc contribution able to avoid detection by the GMM (see Fig.~\ref{fig:residual}). Synthetically adding a thick disc population with varying strength in the VMP regime, we constrain the thick disc contribution in our sample to be rather minor, at $<3\%$. Our sample is less well-suited to identify the presence of a (minor) thin disc component due to our limited coverage of the low-|z| region.

We compare the properties of our sample with those in the literature, especially the works reporting the presence of a disc-like/planar population of stars, using quantities such as $e$, $z_{max}$, $J_\phi$($L_z$), $J_z$ and circularity. We successfully reproduce results of \cite{Sestito_2020}, showing that there exists a large fraction of very metal-poor stars with $z_{max}<3$~kpc as well as an asymmetry between the prograde and retrograde disc-like orbits (see Fig.~\ref{fig:action_action_space}). We conclude that the prograde halo in the very metal-poor regime could be responsible for such an asymmetry. We also generate a sample of mock halo stars from an isotropic NFW distribution function and argue that there is a non-negligible fraction of stars with disc-like orbits (low $e$ and $z_{max}$, or high circularity) in a normal halo population (see Fig.~\ref{fig:e-distribution}). Therefore a combination of cuts on eccentricity and maximum height above the plane and/or the use of circularity is not sufficient for identifying the presence of a disc population and cannot be used to select disc stars cleanly.

There is an overdensity of prograde very metal-poor stars in our sample, as well as in many literature samples at low metallicity, but we show that the presence of a disc population (with large $v/\sigma$) is not needed to explain this. The prograde halo component in our GMM analysis is the main culprit for the overdensity of prograde very metal-poor stars. This supports the conclusions from simulations \citep{Sestito_2021, Santistevan_2021, Belokurov_2022, McCluskey2023, Semenov_2023,Dillamore_2023spin} that most very metal-poor stars were not born in the disc but originate from the time before the disc formed and/or were accreted later on. 

Future large spectroscopic surveys like WEAVE \citep{WEAVE} and 4MOST \citep{4MOST} will collect millions of spectra for stars in the Milky Way, many of those in the Galactic halo. These kind of large, homogeneous samples for low-metallicity stars are necessary to detect possible (statistical) differences between very metal-poor populations formed in different environments. These surveys will hopefully shed more light on the nature of the prograde planar stars in our Galaxy.

\section*{Data availability}

All data used in this work is publicly available. 

\section*{Acknowledgements}
We thank the reviewers for their helpful comments. We thank Federico Sestito for helpful comments on a draft of this work, and Else Starkenburg for suggesting the test in Appendix~\ref{Appendix::B}. HZ thanks the Science and Technology Facilities Council (STFC) for a PhD studentship. AAA acknowledges support from the Herchel Smith Fellowship at the University of Cambridge and a Fitzwilliam College research fellowship supported by the Isaac Newton Trust. VB acknowledges support from the Leverhulme Research Project Grant RPG-2021-205: "The Faint Universe Made Visible with Machine Learning".

This work has made use of data from the European Space Agency (ESA) mission
{\it Gaia} (\url{https://www.cosmos.esa.int/gaia}), processed by the {\it Gaia}
Data Processing and Analysis Consortium (DPAC,
\url{https://www.cosmos.esa.int/web/gaia/dpac/consortium}). Funding for the DPAC
has been provided by national institutions, in particular the institutions
participating in the {\it Gaia} Multilateral Agreement.

\bibliographystyle{mnras}
\bibliography{bibliography}

\appendix

\section{Metal-rich star contamination in the full A23 catalogue}\label{Appendix::0}
\begin{figure*}
    \centering
    \includegraphics[width = \textwidth]{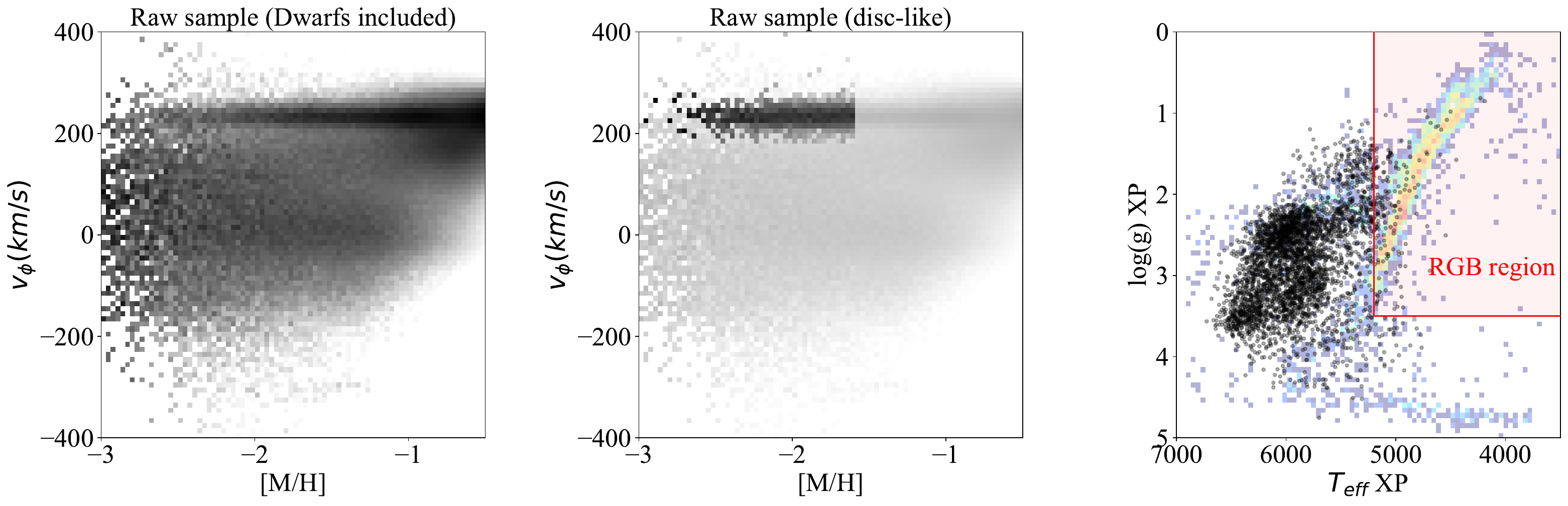}
    \caption{
    In the left panel, we show the column normalised $\mathrm{2d}$ histogram of stars with $\mathrm{fpu}<0.1$ in the full sample from A23, in which a clear excess of high-rotation stars in the metal-poor regime is seen. We highlight the selected metal-poor, thin disc-like stars in the middle panel and plot these stars on the Kiel diagram in the right panel. The background distribution is made by metal-poor stars in APOGEE DR17. Most stars in the selected sample have relatively high temperatures. The red-shaded region is the RGB sample selection criteria in \citet{Andrae_2023}, which removed all these metal-poor, disc-like stars.}
    \label{fig:appendix_kiel}
\end{figure*}

For a reliable chemo-kinematic study of the Milky Way, we only used the vetted RGB sample from \cite{Andrae_2023}, for which the metallicity estimates from the \textit{Gaia}-XP spectra are more reliable. However, in the full A23 catalogue, there appears to be a significant population of metal-poor stars with thin disc-like orbits, as shown in the left panel of Fig.~\ref{fig:appendix_kiel}. Here we investigate the nature of these stars. 

We select metal-poor stars, $\mathrm{[M/H]}<-1.6$, with close-to-circular obits: $v_\phi>180$~km/s, $|v_R|<15$~km/s, and $|v_z|<15$~km/s. The selected stars are highlighted in the middle panel of Fig.~\ref{fig:appendix_kiel}. We present the Kiel diagram for the selected stars in the right-hand panel (black dots), using $T_\mathrm{eff}$, and $\log{g}$ measured from A23. For comparison, we also show the Kiel diagram for stars in the same metallicity range in APOGEE DR17 \citep{APOGEE_DR17}. The metal-poor, disc-like stars are mostly relatively hot ($>5500$~K, where the A23 estimates are less reliable, see \citealt{Andrae_2023}) and occupy an uncommon region in the Kiel diagram between the blue horizontal branch (BHB) and the main sequence (MS). Some blue stragglers (BS) can end up there, but for these stars, the quality of stellar parameters measured using \textit{Gaia}-XP needs to be tested. These are already good reasons to exclude these stars from our analysis. 

\begin{figure}
    \centering
    \includegraphics[width = 0.9\columnwidth]{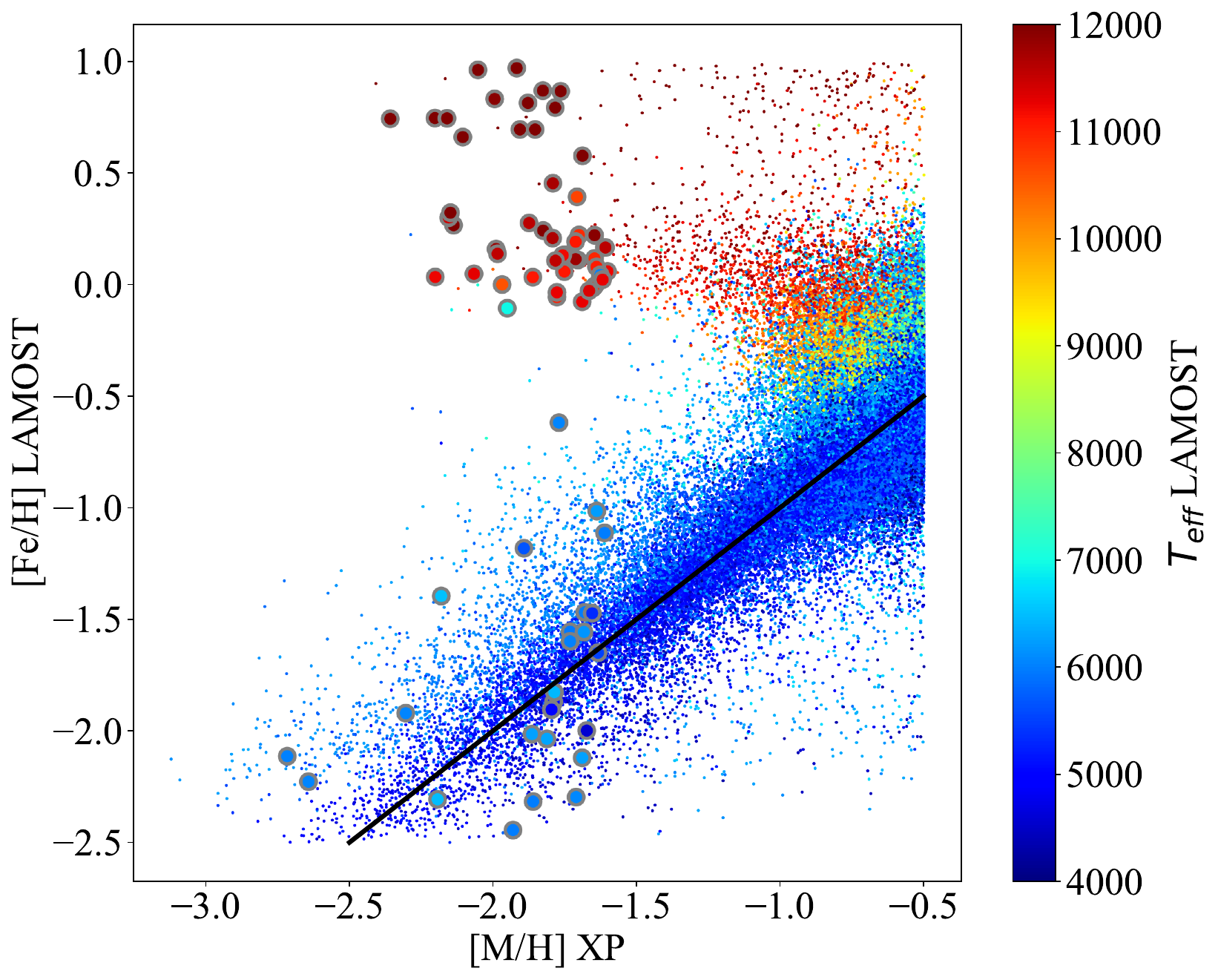}
    \caption{A comparison between the A23 \textit{Gaia}-XP metallicity (x-axis) and LAMOST metallicity (y-axis) for stars in common between LAMOST and A23 (with  $[M/H]_{XP}<-0.5$). The colour shows the effective temperature measured by LAMOST for each star. The large circles are selected metal-poor, disc-like stars (see the text and Figure~\ref{fig:appendix_kiel}), which are apparently mostly hot, metal-rich stars according to LAMOST. The black-solid line represents the 1:1 line. 
    }
    \label{fig:appendix_lamost}
\end{figure}

We further verify the quality of the stellar parameters for these metal-poor, disc-like stars by cross-matching the full A23 catalogue with the 8th data release from the low-resolution spectroscopic LAMOST survey \citep{Deng_2012_LAMOST}. 
In Fig.~\ref{fig:appendix_lamost}, we compare the metallicity measured from \textit{Gaia}-XP and LAMOST spectroscopy with the colour representing the effective temperature measured by LAMOST. The small dots in the background are all stars with \textit{Gaia}-XP metallicity below $\mathrm{[M/H]_{XP}}<-0.5$. There are clouds of hot ($>10\,000$~K) metal-rich stars in LAMOST that are assigned low metallicity in the A23 catalogue. The large circles are for the selected metal-poor (from A23), disc-like stars, showing that many of these reside in the contamination cloud. We have checked the LAMOST spectra of a random subset of these stars, and they are indeed very hot stars. We suspect that because the metal lines in these stars are small, the analysis of A23 instead assigns them a low metallicity (their training sample does not include stars beyond 7000~K). 

In summary, there are good reasons to stick to cooler RGB stars instead of using the full A23 sample. This supports our decision to perform our rigorous analysis on the vetted giants sample from A23 only.

\section{Full GMM fitting plots}\label{Appendix::A}
\begin{figure*}
    \centering
    \includegraphics[width = \textwidth]{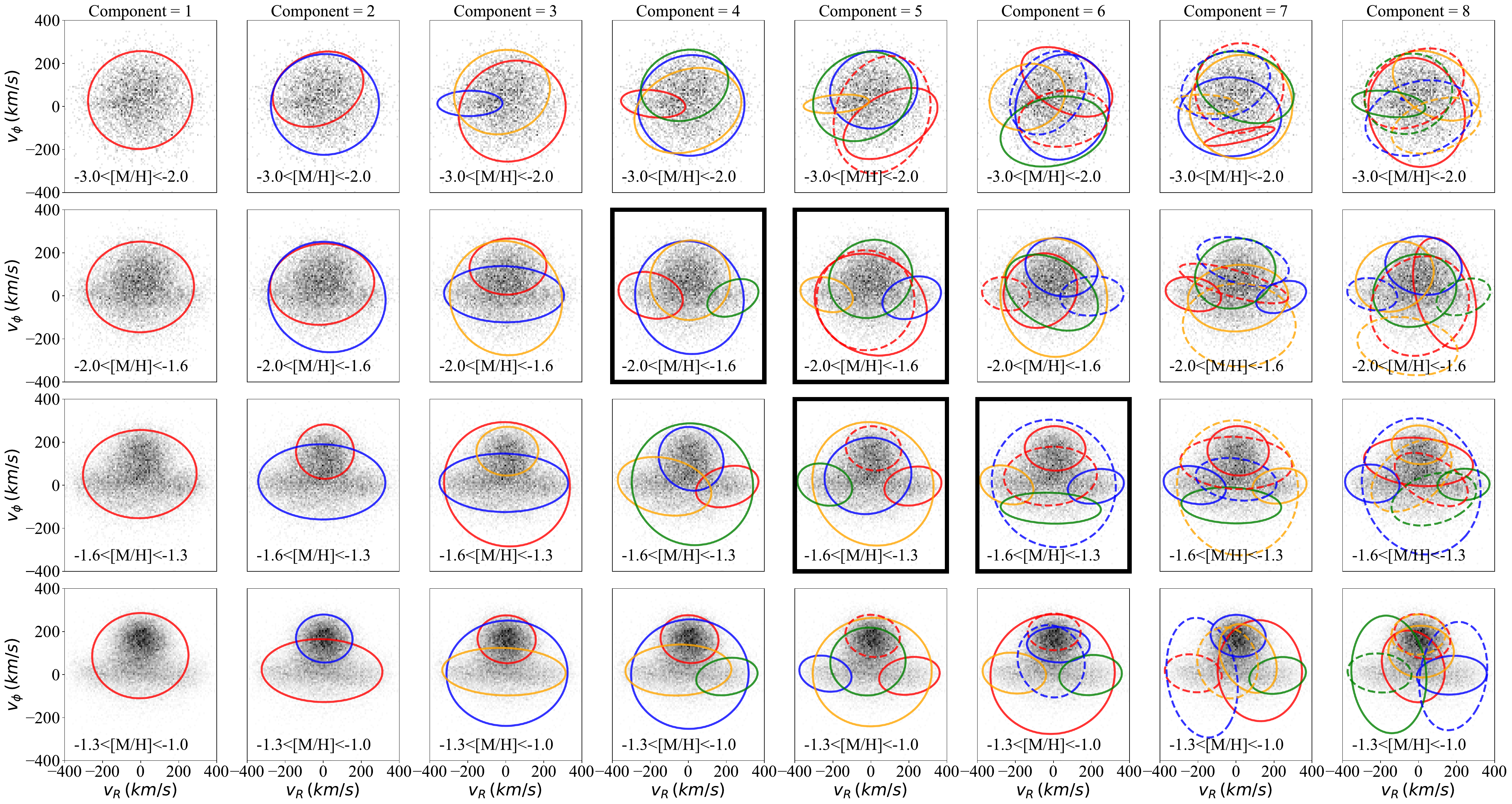}
    \caption{This figure shows all the best-fit GMMs results for each $N$ in each metallicity bin, for the VMP (top row), IMP (second row), MP2 (third row) and MP1 (bottom row) selections. Each subplot corresponds to a point on the black BIC line in Fig.~\ref{fig:BIC}. The different ellipses with different colours and line styles represent the 2$\sigma$ contour of the Gaussian component identified. We highlight the controversial GMM fitting with close BIC values with a bold subplot frame.}
    \label{fig:appendix}
\end{figure*}

We fit the sample with GMMs in the ($v_R$, $v_\phi$, and $v_z$) space with $N$-components in each metallicity bin. For each $N$, we run 50 trials with different initial guesses to seek the best-fitted GMM in the global minimum, and the best-fitted GMMs for each $N$ are shown in Fig.~\ref{fig:appendix}. We use the BIC value as an indication to choose the optimised $N$, avoiding over-fitting. The trends of the BIC values as a function of $N$ in each metallicity bin are shown in Fig.~\ref{fig:BIC}. Also mentioned in Section~\ref{sec::model}, around the minimum of the BIC values, some $N$-component GMM fitting share similar BIC values (e.g. 4/5-component models in the IMP bin, $-2<\mathrm{[M/H]}<-1.6$; 5/6-component models in the MP1 bin, $-1.6<\mathrm{[M/H]}<-1.3$). In these cases, the BIC value can no longer differentiate the optimised $N$, and we always prefer the lower $N$ because it is easier to interpret the GMM physically. 

The subplots associated with these GMMs are highlighted with bold frames in Fig.~\ref{fig:appendix}. In the IMP bin, the prograde halo, and two GS/E components are common in both 4 and 5-component GMMs. However, the 5-component GMM model shows two stationary halos with similar kinematic properties. Due to the redundancy of the extra halo component, we prefer the 4-component GMM in the IMP bin. Adding even more components to fit stars in the IMP bins, the GMM does not identify sub-populations with disc-like kinematics, strengthening our arguments that there is no disc component in the IMP regime. In the MP2 bin, the 5 and 6-component models have a similar BIC value. The 6-component GMM identifies an extra retrograde halo component with a significantly larger velocity dispersion in the radial direction than the azimuthal direction. However, as the purpose of the paper is not to decipher the detailed halo composition, we do not dive further to explore the property of this retrograde halo structure, and adopt the 5-component model as our preferred one. A disc component is ubiquitously found in all MP2 GMMs with $N\geq2$, which demonstrates that the disc stars to emerge between $-1.6<\mathrm{[M/H]}<-1.3$. 

The BIC trends in the VMP and MP2 bins are less controversial. We emphasised in the section~\ref{sec::vmpsubstructure} the overdensity of stars with negative $v_R$, high eccentricity, and little net rotation. In the top row of Fig.~\ref{fig:appendix} it is shown in the blue ellipse on the 3-component VMP model, and a similar overdensity exists in the higher $N$ GMMs as well. It is worth investigating the origin of this overdensity in the future. We note that even in the higher $N$ GMM models for the VMP selection, no disc component is identified. 

\section{Evolution of $\mathbf{v_R - v_\phi}$ when matching the VMP spatial distribution}\label{Appendix::B}

\begin{figure*}
    \centering
    \includegraphics[width = \textwidth]{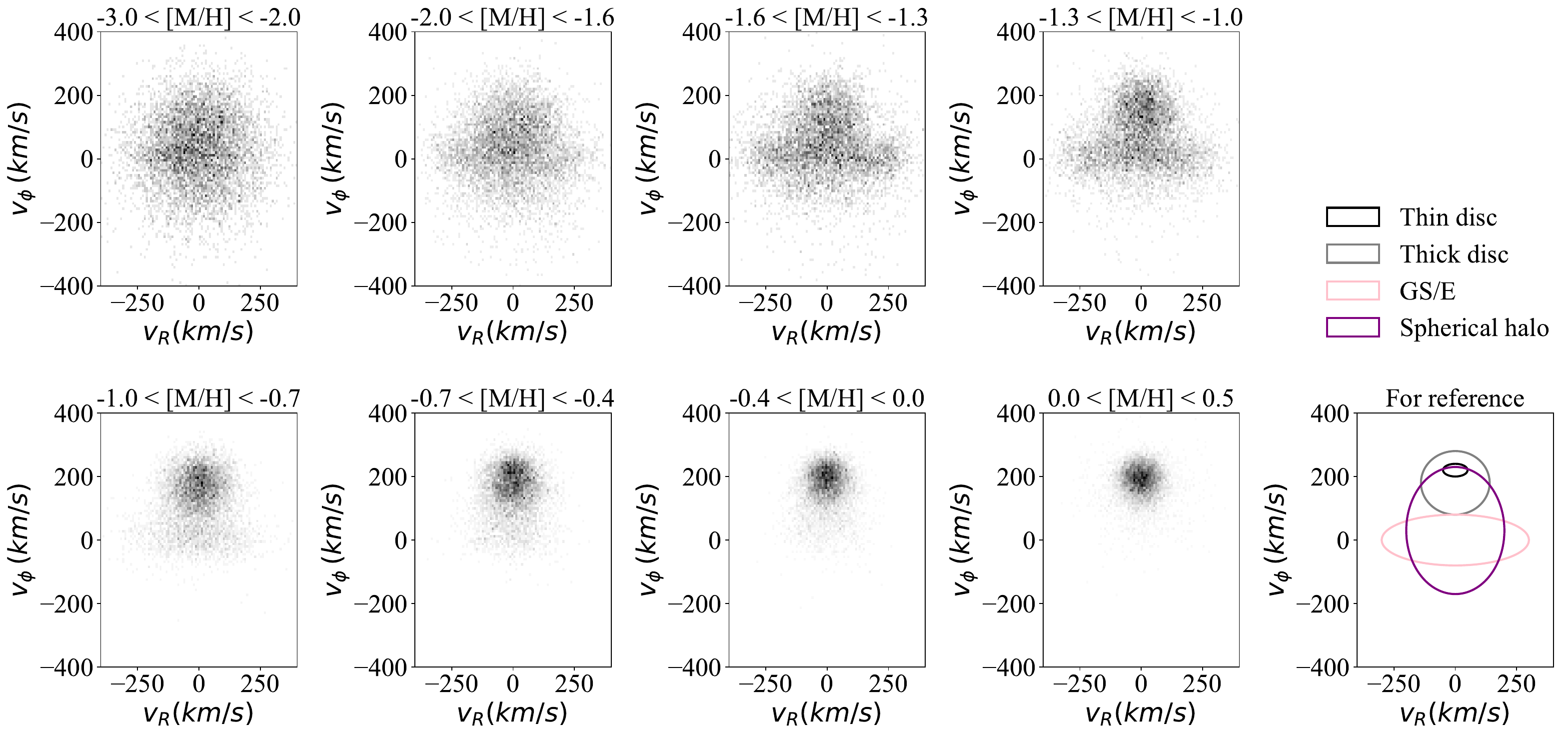}
    \caption{This figure is the same as Fig.~\ref{fig:vr_vphi_space}, but now the spatial distribution of stars in each bin matches the $R-z$ distribution for the VMP bin (see text). The number of stars is the same in each panel.}
    \label{fig:appendixB}
\end{figure*}

Even though the quality cuts we applied to all metallicity bins are the same, the spatial distribution still slightly differs for each bin. The metal-rich populations ([M/H]~$>-1.0$) are more concentrated to the solar neighbourhood (hence, closer to the galactic disc). Therefore, there is a greater chance of detecting a disc population among the metal-rich stars than the metal-poor stars. To test the significance of this effect, we force the $R-z$ distribution of all metallicity bins to align with the $R-z$ distribution of the VMP bin. For every VMP star in our sample, and for each metallicity bin, we find the star matching closest in the $R-z$ plane. This results in a sample of 6800 stars (the total number of VMP stars in our sample) in each bin reproducing the $R-z$ distribution of the VMP bin. We present the corresponding $v_R-v_\phi$ distribution in Fig.~\ref{fig:appendixB}, to compare with Fig.~\ref{fig:vr_vphi_space} in the main manuscript. Although the metal-rich population is now slightly less concentrated to the disc region, the disc population is still clear and qualitatively the differences between Fig.~\ref{fig:appendixB} and Fig.~\ref{fig:vr_vphi_space} are small. We still see a smooth transition from halo-dominated metal-poor stars to the disc-dominated metal-rich stars. Hence, if a significant disc-population exists in the VMP regime, we should be able to detect it. This test shows that the difference in the spatial distribution in each metallicity bin is a minor issue and should not affect our arguments regarding the existence of a VMP disc population.

\bsp    
\label{lastpage}
\end{document}